\documentclass{aa}  

\usepackage{graphicx}
\usepackage{amsbsy}
\usepackage{txfonts}
\usepackage{hyperref}
\hypersetup{
       hyperfootnotes=true,                     
           bookmarks=true,                      
           colorlinks=true,
           linkcolor=black,
            linktoc=page,
           anchorcolor=black,
           citecolor=blue,
           urlcolor=black,
}
\DeclareUnicodeCharacter{2212}{-}
\usepackage[dvipsnames,svgnames,x11names]{xcolor}

\begin{document} 

    \title{Cooling rate and turbulence in the intracluster medium of the cool-core cluster Abell 2667}


   \author{M. Lepore
          \inst{1,2}
           \and
          C. Pinto
          \inst{3}
          \and
          P. Tozzi
          \inst{1}
          \and
          M. Gaspari
          \inst{4}
          \and
          F. Gastaldello
          \inst{5}
          \and
          A. Liu
          \inst{6,1,7}
          \and
          P. Rosati
          \inst{8}
          \and
          R. van Weeren
          \inst{9}
          \and
          G. Cresci
          \inst{1}
          \and
          E. Iani
          \inst{10}
          \and
          G. Rodighiero
          \inst{11,12}
        }

\institute{INAF-Osservatorio Astrofisico di Arcetri, Largo Enrico Fermi 5, 50125, Firenze, Italy
\and Dipartimento di Fisica e Astronomia, Università degli Studi di Firenze, Via Sansone, 1 - 50019 Sesto Fiorentino, Firenze, Italy
\and INAF – IASF Palermo, Via Ugo La Malfa 153, 80146, Palermo, Italy
\and Department of Physics, Informatics and Mathematics, University of Modena and Reggio Emilia, 41125 Modena, Italy
\and INAF-IASF Milano, Via A. Corti 12, I-20133 Milano, Italy
\and Max Planck Institute for Extraterrestrial Physics, Giessenbachstrasse 1, 85748 Garching, Germany
\and Institute for Frontiers in Astronomy and Astrophysics, Beijing Normal University, Beijing 102206, China
\and Dipartimento di Fisica e Scienze della Terra, Università degli Studi di Ferrara, via Saragat 1, I-44122 Ferrara, Italy
\and University of Leiden, Rapenburg 70, 2311 EZ Leiden, The Netherlands
\and Kapteyn Astronomical Institute, University of Groningen, P.O. Box 800, 9700 AV Groningen, The Netherlands
\and Department of Physics and Astronomy, Università degli Studi di Padova, Vicolo dell’Osservatorio 3, I-35122, Padova, Italy
\and INAF - Osservatorio Astronomico di Padova, Vicolo dell’Osservatorio 5, I-35122, Padova
}

\abstract
   {We present a detailed analysis of the thermal X-ray emission from the intracluster medium 
   in the cool-core galaxy cluster Abell 2667 at $z=0.23$. }
   {Our main goal is to detect low-temperature ($<2$ keV) X-ray emitting gas associated with a potential 
   cooling flow connecting the hot intracluster medium reservoir to the cold gas phase responsible for 
   star formation and supermassive black hole feeding.}
   {We combined new deep \textit{XMM-Newton} EPIC and RGS data, along with archival \emph{Chandra} data, and performed a spectral analysis of the emission from the core region.}
   {We find 1$\sigma$ upper limits on the fraction of gas cooling equal to $\sim$ 40 $\rm M_{\odot}yr^{-1}$ and $\sim$ 50-60 $\rm M_{\odot}yr^{-1}$, in the temperature ranges of 0.5-1 keV and 1-2 keV, respectively. We do not identify OVII, FeXXI-FeXXII, and FeXVII recombination and resonant emission lines in our RGS spectra, implying that the fraction of gas cooling below 1 keV is limited to a few tens of solar masses per year at maximum. 
   We do detect several lines (particularly SiXIV, MgXII, FeXXIII/FeXXIV, NeX, OVIII$\alpha$) from
   which we are able to estimate the turbulent broadening. We obtain a 1$\sigma$ upper limit of 
   $\sim$ 320 km/s, which is much higher than the one found in other cool-core clusters such as Abell~1835, 
   suggesting the presence of some mechanisms that boost significant turbulence in 
   the atmosphere of Abell~2667. Imaging analysis of {\sl Chandra} data suggests the presence of
   a cold front possibly associated with sloshing or with intracluster medium cavities.  However, current data do 
   not allow us to clearly identify the dominant physical mechanism responsible for turbulence. }
  {These findings show that Abell 2667 is not different from 
  other, low-redshift, cool-core clusters, with only upper limits on the mass deposition rate associated 
  with possible isobaric cooling flows. Despite the lack of clear signatures of recent feedback events, the large upper limit on the turbulent velocity leaves room for significant 
  heating of the intracluster medium, which may quench cooling in the cool core for an extended period, albeit also driving local intracluster medium fluctuations that will contribute to the next cycle of condensation rain.}


   \maketitle

\section{Introduction}
\label{sec:Introduction}
Galaxy clusters are the largest gravitationally bound structures in the Universe. 
These are formed by baryonic and nonbaryonic matter in different states 
and with different assembly histories. Typically they host 100-1000 galaxies 
in virial equilibrium with a total halo mass varying between $10^{14}$ M$_{\odot}$ and 
$10^{15}$ M$_{\odot}$ and a diameter of a few megaparsecs. 
The stellar mass in member galaxies 
typically amounts to  2-5$\%$ only of the total observed mass. The majority of the
mass of the halo is contributed by dark matter, with a mass fraction of $\sim$ 0.8.
Most of the remaining mass budget (15-18$\%$) 
is contributed by a tenuous, high-temperature ($\sim$ $10^7-10^{8}$ K) plasma 
with a typical electron density on the order of $10^{-3}$ cm$^{-3}$: 
the intracluster medium (ICM). The ICM is expected to be in an approximate 
hydrostatic equilibrium in the cluster potential well and can be detected 
in the X-ray band thanks to its thermal bremsstrahlung emission plus
line emission from heavy element ions.

In some clusters, the ICM density profile can be described by a single 
$\beta$ model \citep{1978Cavaliere} that 
is rapidly decreasing with 
radius toward the outskirts outside a flat core.  This description, however, 
is not accurate for many clusters that show sharply peaked profiles, 
dubbed cool cores, which can reach electron densities as high as 0.1 cm$^{-3}$
in the center.  In these cases, a double $\beta$ model is required, 
suggesting that physical processes acting in the core are modifying 
the ICM distribution. 
This is not surprising, since the gas cooling time, which scales as the inverse of the density, is measured to be much shorter than the Hubble time in cool cores 
\citep{2005McNamara, 2006Fabian, 2014Vantyghem, 2014Sanders} over a fairly large redshift interval \citep{2010Santos}. 

This initially led to the 
conclusion that a massive cooling flow (CF) is developing in the ICM 
of most clusters with cool cores \citep{1977Fabian}. According 
to the original isobaric CF scenario \citep{1994Fabian}, the gas 
in the central regions of clusters should experience adiabatic 
compression at constant pressure, cooling mainly through X-ray emission, and
inevitably condensate into clouds of cold gas. 
This picture is supported by the 
observation of temperature drops toward the center 
\citep{2000McNamara, 2001Tamura, 2007Sanders, 2009Cavagnolo, 2024Sonkamble, 2024Liu}.  
In addition, the emission measure of the ICM as a function of temperature 
can be accurately predicted in the framework of the adiabatic CF model, 
and therefore the associated mass deposition rates (MDRs) can be directly inferred by 
the spectral distribution of the X-ray emission from the cool core. 
The MDR values based 
on the isobaric CF model and estimated from the total core luminosity are typically estimated to be 
in the range of 100-1000 $\rm M_{\odot}$\,yr$^{-1}$. 
However, the bulk of 
the X-ray emission is contributed by the relatively hot ICM, while the contribution
from the coldest, X-ray emitting gas amounts to a few percent of the total
luminosity. Typically, the cooling gas is not directly observed in CCD spectra.

Since the ICM is enriched with heavy ions, whose abundance is even higher toward the 
center, another key prediction of the isobaric CF model 
is the presence of a line-rich X-ray spectrum due 
to the presence of many transitions in the ionized heavy elements, which can be observed in
high-resolution X-ray grating spectroscopy. 
Surprisingly, the analysis of the first high-resolution X-ray spectra, mostly thanks to \textit{XMM-Newton} observations, clearly ruled out the isobaric CF model (the so-called soft CF problem; see \citealp{2001Peterson,2001MolendiPizzolato,2001Kaastra,2002Xu, 2002Ettori}), 
due to the lack of emission lines associated with gas colder than 
about one third of the virial value \citep{2004Donahue} in any cool-core clusters. 
Since it is extremely unlikely that the coldest gas is systematically poor in metals, 
the lack of emission lines directly implies the lack of the low-temperature gas predicted by the isobaric cooling model.  

Given the quality of current high-resolution X-ray spectra, 
the presence of CFs is not completely ruled out, but they are
constrained to be more than one order of magnitude lower than the ones 
predicted in the isobaric CF model. In addition, the maximum MDRs
allowed by current data are at least one order of magnitude lower than the star formation 
rates (SFRs) observed in the hosted brightest cluster galaxy 
(BCG; \citealp{2016Molendi}), implying a significant difference
between the timescale of cold gas consumption and the timescale of gas cooling. 
To date, the only convincing CF candidate is the 
well-documented Phoenix cluster at $z\sim 0.6$ 
(see \citealp{2012McDonald, 2019McDonald, 2015Tozzi, 2018Pinto}). 
This cluster exhibits a starburst of 500-800 $\rm M_{\odot} yr^{-1}$ 
\citep{2015McDonald, 2017Mittal}, comparable to a cooling rate of 
$\dot M=350^{+250}_{-200}$ $\rm M_{\odot} yr^{-1}$ below 2 keV (uncertainties at a 90\% confidence level)
measured with \textit{XMM-Newton} Reflection Grating Spectrometer (RGS) data 
(\citealp{2018Pinto}; see also \citealp{2015Tozzi}, which found 
$\dot M = 120^{+340}_{-120}$ $\rm M_{\odot} yr^{-1}$ below 3 keV from RGS data).

Given the difficulty in measuring MDRs, it is not possible
to firmly identify the clusters of galaxies actually hosting a CF. 
Phenomenologically, galaxy clusters are classified as cool-core 
(CC; relaxed systems with a peak in the surface brightness and mostly with regular 
and spheroidal shapes) and non-cool-core (NCC; systems with no bright cores typically 
with disturbed morphologies likely associated with ongoing merging processes;
\citealt{2001MolendiPizzolato, 2010Hudson, 2016Zhang}). On a general basis, CFs are expected to be hosted only by CC clusters, but their occurrence and magnitude are not quantified, and we do not have information about their duty cycle.

The inadequacy of the isobaric CF model inevitably raises the question 
of the mechanism preventing the gas from cooling in adiabatic mode. 
As of today, there is a large consensus on the heating from 
mechanical feedback and turbulence, mostly provided by  
the central active galactic nucleus (AGN; e.g.,
\citealp{2007McNamara,2012Fabian,2022McKinley}). 
High-resolution hydrodynamical simulations show that AGN jets or outflows can inject 
a sufficient amount of energy into the ICM to offset both the global cooling rates 
\citep{2012Gaspari} and the differential luminosity distribution toward the 
soft X-ray via nonisobaric modes \citep{2015Gaspari}. The AGN jet feedback 
is also crucial to stir up the ICM “weather” with substantial enstrophy 
(magnitude of vorticity) in the cluster cores (\citealt{2020Wittor,2023Wittor}).
This feedback mechanism has been directly observed in many systems, such as nearby 
clusters, Perseus \citep{2003Fabian, 2011Fabian}, Hydra A 
\citep{2000McNamara, 2022Timmerman}, Virgo \citep{2012Werner, 2015Russell}, 
and Abell 2626 \citep{2019Kadam}.  On the other hand, residual star formation 
in the BCG, the presence of cold and multiphase gas 
surrounding BCGs observed with ALMA (e.g., \citealp{2021North}) and MUSE 
(e.g., \citealp{2021Maccagni,2022Olivares}), and the distribution of heavy elements
in the hot gas (e.g., \citealt{2019LiuA,2020LiuA,2021Gastaldello}), 
suggest that at least some fraction of the hot ICM underwent cooling at some 
point during the secular evolution of clusters.  The key point is that, 
to agree with current observations, CFs should be 
hidden, characterized by very low MDRs, 
or occurring on short timescales. Some recent works show that a substantial 
fraction of the cooling gas may be intrinsically absorbed (\citealp{2022Fabian}), 
interpreting the observed far-infrared (FIR) luminosity from the core as a consequence 
of the cooling rate whose emission is reprocessed by the surrounding medium.  
In addition, mechanical AGN feedback can stimulate the condensation of 
the hot phase via turbulent nonlinear thermal instability, leading to a rain of 
clouds or filaments 
that are expected to boost the SMBH feeding rate via 
chaotic cold accretion (CCA; \citealt{2020Gaspari}, for a review), characterized by 
a highly variable MDR. 
Finally, massive, short-lived CFs may occur and replenish the cold gas reservoir 
before being quenched by feedback processes from the central SMBH in the BCG. 
In every scenario, the central AGN is expected to play a key role in regulating 
the thermodynamics in the ICM, while shaping 
the hot halo properties \citep{2021Pasini} and SMBH scaling relations \citep{2019Gaspari}.

One of the crucial open questions  
is how much gas cools and trickles below 1.0 keV, down to the lowest temperatures 
detectable in the X-ray grating spectra ($0.3$\,-\,$0.5$ keV).
In this work, we reconsider this question using X-ray images and spectra taken with 
XMM European Photon Imaging Camera (EPIC), XMM/RGS and \emph{Chandra} Advanced CCD Imaging Spectrometer 
(ACIS) of the CC cluster Abell 2667 (hereafter A2667). This cluster 
has a BCG with prominent ongoing activity with peculiar features 
such as blue and red wings, 
broadening of the lines, and offsets with respect 
to the BCG reference system, suggesting the presence of multiple gas
components on scales of a few tens of kiloparsecs. 
It also shows a prominent AGN signature 
in the optical, radio, and X-ray bands, with a strongly absorbed 
X-ray nuclear emission ($\sim$ 3 $\times$ $10^{43}$ erg $\rm s^{-1}$), 
and diffuse ICM emission on the order of 3 $\times$ $10^{45}$ erg $\rm s^{-1}$, 
comparable to that of the Phoenix cluster \citep{2012McDonald}. 
The similarity of the environmental 
properties with the Phoenix is at variance with the large difference 
in the strength of the activity: the nuclear emission is $\sim$ 100 and 
20 times lower in A2667 in the X-ray and radio band \citep{2014vanWeeren},
respectively, and the total SFR in the BCG is a factor of $\sim$ 50 lower.  
This composite picture makes the BCG of A2667 an interesting 
candidate in which to study the cooling and heating processes in clusters.

This paper is structured as follows. In Sect. \ref{sec:A2667}, 
we describe the previous results for A2667 available in the literature. 
In Sect. \ref{sec:chandra}, 
we describe the archival \emph{Chandra} data. 
In Sect. \ref{sec:data}, we describe the new XMM-EPIC and RGS 
deep observation used in this work. In Sects. \ref{sec:spec EPIC}, 
\ref{sec:spec RGS}, and \ref{sec:spec_joint} 
we describe XMM-EPIC and XMM-RGS data analysis and results. 
In Sect. \ref{sec: discussion}, we discuss the results obtained from our analysis, 
while in Sect. \ref{sec:future perspectives} we comment on the possible extension of 
this study with future X-ray facilities. Finally, our conclusions are summarized in Sect. \ref{sec:Conclusions}. Throughout this paper, we adopt the current 
Planck cosmology with $\rm \Omega_{m}$=0.315, and 
$\rm H_{0}$=67.4 km $\rm s^{−1}$ $\rm Mpc^{−1}$ in a flat geometry \citep{2020Planck}. 
In this cosmology, at $z=0.23$, 1 arcsec corresponds to 3.81 kpc, 
and the Universe is 11.01 Gyr old. Quoted error bars correspond to a $1\sigma$ 
confidence level, unless otherwise stated.

\section{A2667: Previous observations and main results}
\label{sec:A2667}

A2667 
is a lensing galaxy cluster 
in the nearby Universe ($z=0.233$). It has multiple image systems, including 
one of the brightest giant gravitational arcs with an elongation of $\sim 10$ arcsec
at $z = 1.0334$ \citep{2006aCovone,2006bCovone}. It is also associated 
with a radio minihalo \citep{2017Giacintucci, 2022Knowles} with a largest angular size of $4.6$' and a largest physical linear size at the cluster redshift of $1.02$ Mpc 
\citep{2022Knowles}. This cluster is among the top 5$\%$ most luminous X-ray 
clusters observed in the ROSAT All Sky Survey, with a luminosity of
$\rm L_{x}  = (13.39 \pm 0.25) \times \, 10^{44}$ erg/s 
in the range of [0.1-2.4] keV  within $\rm r_{500}=1.38\pm0.07$ Mpc \citep{2016Mantz}. 
It is classified as a relaxed cluster \citep{2009Gilmour, 2015Kale, 2012Mann} 
with a BCG at the center (RA=23:51:39.40 Dec=−26:05:03.3). 
The BCG hosts an X-ray AGN with intrinsic absorption 
$\rm N_{H}(10^{22} cm^{-2}$)=$13.7^{+7.6}_{-4.9}$ 
and X-ray luminosity $\rm log(L_{X[2-10] keV})[erg/s]=43.4^{+0.2}_{-0.4}$ \citep{2018Yang}. 
The BCG has a stellar mass of $\rm log(M_{\rm bulge}/M_{\rm \odot})$=11.97$\pm$0.04 
and hosts a SMBH whose mass is estimated to be $\rm log(M_{BH}/M_{\odot})$=10.35$\pm$0.27
\citep{2019Phipps} from the standard fundamental plane relation \citep{2003Merloni}. 

The SFR in the BCG has been estimated to range between 
SFR=$8.7\pm 0.2$ $\rm M_{\odot}$ $\rm yr^{-1}$ (via the spectral energy
distribution from the FIR \emph{Spitzer} and \emph{Herschel} data; \citealp{2012Rawle})
and $\simeq$ 55 $\rm M_{\odot}$ $\rm yr^{-1}$  (via MUSE spectra; \citealp{2019Iani}). 
The difference between the two estimates is a consequence of the methodologies applied and they can be considered as  
lower and upper limits, respectively. In effect, while the value inferred 
from the FIR does not take into account the contribution from 
unobscured star formation, the estimate by \cite{2019Iani} may be 
contaminated by the AGN emission. The star formation is responsible for
previously noticed strong $H\alpha$ and $\rm [OII]\lambda 3727 \AA$ lines 
\citep{1998Rizza}, which are often associated with CCs.

The bidimensional modeling of the BCG optical surface brightness profile reveals 
the presence of a complex system of substructures extending all around the galaxy. 
Clumps of different sizes and shapes plunged into a more diffuse component constitute 
these substructures, whose intense “blue” 
optical color hints at the presence 
of a young stellar population \citep{2019Iani}. However, A2667's central region 
seems to be dominated by an evolved and passively aging stellar population \citep{2006aCovone}.

In the radio band, \cite{2015Kale} classified this central galaxy as a
radio-loud AGN with a radio power of $P=3.16 \times 10^{24}$ W $\rm Hz^{−1}$, 
using the National Radio Astronomy Observatory (NRAO) VLA Sky Survey radio data at 1.4 GHz.
The X-ray morphology \citep{1998Rizza} and dynamical analysis \citep{2006aCovone} 
support the idea that A2667's inner core is relaxed and shows a drop 
in the ICM temperature profile in the central 100 kpc, suggesting that the cluster hosts 
a CC with an average cooling time of $\sim$ 0.5 Gyr \citep{2009Cavagnolo}. 
 Indeed, \cite{2016Zhang} classified this luminous cluster as a strong CC system with $\rm r_{cool}(kpc)$=135.9$^{+12.9}_{-14.2}$.

\section{Chandra: Observations and data reduction}
\label{sec:chandra}

Before proceeding with the reduction and analysis of our deep \textit{XMM-Newton} observation, 
we revisited the shallow {\sl Chandra} data available on the archive, provided with 
the highest spatial resolution, in particular to compute a \emph{Chandra} 
soft band image (0.5-2 keV) and compute the surface brightness profile, which is 
necessary to model the (instrumental) line broadening expected in the 
RGS spectra. 

A2667 was observed for 10 ks with ACIS-S granted in Cycle 02 (PI: S. Allen). 
The observations were completed in June 2001 with a single pointing. 
We performed data reduction starting from the level = 1 event files with 
\texttt{CIAO 4.13}, with the latest release of the \emph{Chandra} Calibration Database 
at the time of writing (\texttt{CALDB 4.9.4}). We ran the task \texttt{destreak} 
to flag and remove spurious events (with moderate to small pulse heights) along single rows.
We ran the tool 
\texttt{acis$\_$detect$\_$afterglow} to flag residual charge from cosmic rays 
in CCD pixels.  Finally, as this observation is taken in the VFAINT mode, 
we ran the task \texttt{acis$\_$process$\_$events} with the parameter 
\texttt{apply$\_$cti=yes} to flag background events that are most likely associated with cosmic rays and reject them. With this procedure, the ACIS particle background can be significantly reduced compared to the standard grade selection. The data were then filtered to include only the standard event grades 0, 2, 3, 4, and 6. The level 2 event files generated in this way were visually inspected for flickering pixels or hot columns left after the standard reduction but we found none after filtering out bad events. We also carefully inspected the image of the removed photons, particularly to verify whether we have pile-up effects\footnote{https://cxc.cfa.harvard.edu/ciao/ahelp/acis$\_$pileup.html}. Finally, time intervals with high background were filtered by performing a 3$\sigma$ clipping of the background level. The light curves were extracted in the 2.3–7.3 keV band and binned with a time interval of 200 s. The time intervals when the background exceeds the average value by 3$\sigma$ were removed with the tool \texttt{deflare}. The final total exposure time after data reduction and excluding the dead-time correction amounts to 9.6 ks (corresponding to the LIVETIME keyword in the header of \emph{Chandra} fits files). 
Images were created in the 0.5-2 keV, 2-7 keV, and total (0.5-7 keV) bands, with no binning to preserve the full angular resolution of $\sim$ 1 arcsec (FWHM) at the aimpoint (1 pixel corresponds to 0.492 arcsec). In the left panel of Fig. \ref{fig:chandra XMM A2667} we show the total \emph{Chandra} image. 


\section{\textit{XMM-Newton}: Observations and data reduction}
\label{sec:data}

\subsection{XMM-Newton: EPIC data}
\label{subsec:EPIC data}

We obtained a total of 274 ks with \textit{XMM-Newton} on A2667 in AO23 (Proposal ID 90028, PI: P. Tozzi). Data were acquired in December and June 2022 (see Table \ref{table:XMM data} for exposure times obtained after data reduction).
To these new observations, we also added to our analysis the archival 
observation ObsID 0148990101 taken in 2003  (Proposal ID 14899, PI: S. Allen).

\begin{figure*}
   \centering
   \includegraphics[scale=0.30]{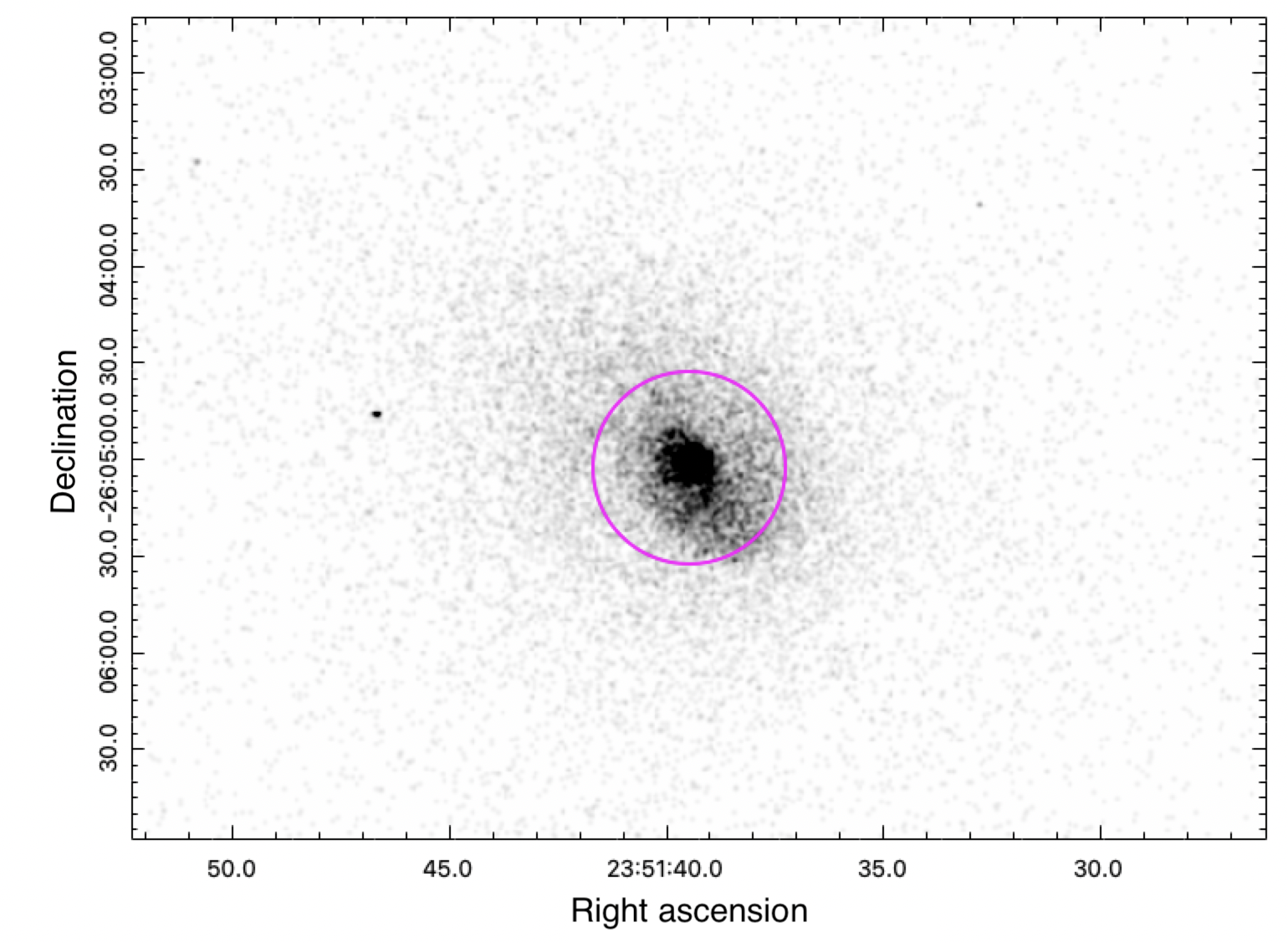}
    \includegraphics[scale=0.32]{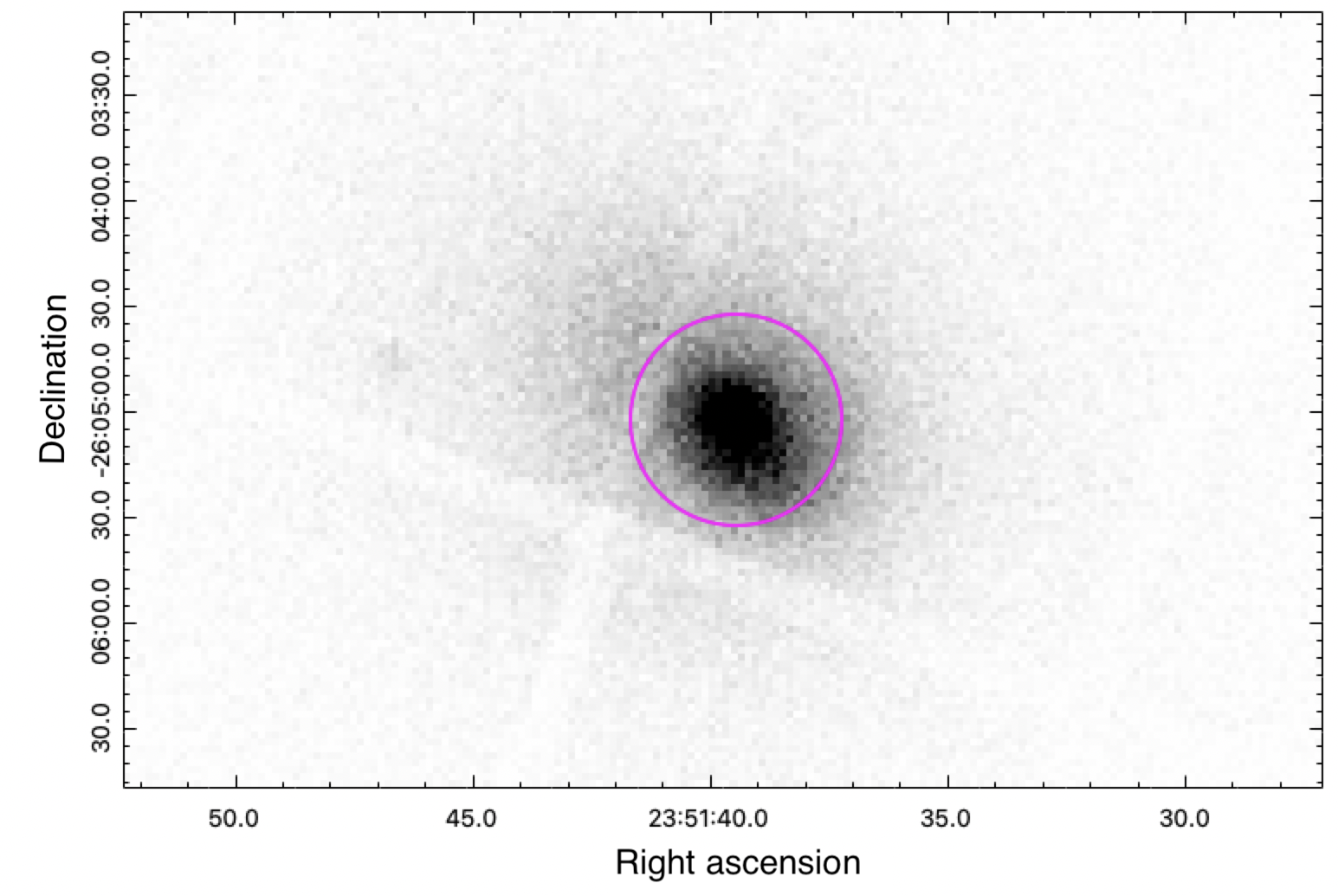}
   \caption{Chandra and XMM images of A2667's central region. 
   \emph{Left panel}: \emph{Chandra} [0.3-10 keV] image of A2667 in a FoV of $\sim$ 6 $\times$ 4 arcmin$^{2}$. \emph{Right panel}: \textit{XMM-Newton}/EPIC MOS1+MOS2+pn [0.5-7 keV] image of A2667 in a FoV of $\sim$ 6 $\times$ 4 arcmin$^{2}$. The magenta circle in both images shows the spectral extraction region of 0.5 arcmin radius used in XMM/EPIC data analysis. }
    \label{fig:chandra XMM A2667}
\end{figure*}

The observation data files (ODFs) were processed to produce
calibrated event files using the most recent release of the
\textit{XMM-Newton} Science Analysis System (SAS v21.0.0), with
the calibration release XMM-CCF-REL-323, and running the
tasks \texttt{epproc} and \texttt{emproc} for the pn and MOS, respectively,
to generate calibrated and concatenated EPIC event lists. Then,
we filtered EPIC event lists for bad pixels, bad columns, cosmic-ray
events outside the field of view (FoV), photons in the gaps
(FLAG=0), and applied standard grade selection, corresponding
to PATTERN < 12 for MOS and PATTERN $\leq$ 4 for pn. We
removed soft proton flares by applying a threshold on the count
rate in the 10-12 keV energy band. To define low-background
intervals, we used the condition RATE $\leq$ 0.35 for MOS and
RATE $\leq$ 0.4 for pn\footnote{see "Users Guide to the \textit{XMM-Newton} Science Analysis System", Issue 18.0, 2023 (ESA: \textit{XMM-Newton} SOC)}. 
For each ObsID, we removed out-of-time (OoT) events from the pn event file and spectra, 
running the command \texttt{epchain} to generate an OoT event list and applying 
the same corrections and selections adopted for pn data reduction.  In the right panel of Fig. 
\ref{fig:chandra XMM A2667}, we show the total \textit{XMM-Newton}/EPIC MOS1+MOS2+pn image in the 0.5-7 keV energy band obtained after data reduction.

To generate the effective area files, we considered the effective area corrections, 
setting the parameter \texttt{applyxcaladjustment=yes} in the arf file generation 
\texttt{arfgen} to reduce spectral differences of the EPIC-MOS 1 and 2 
with respect to EPIC-pn detectors above 2 keV. After a first comparison 
of the spectra with both corrections, we considered only the effective area correction, 
since the one for OoTs is negligible. Finally, effective area and response matrix files 
were generated with the tasks \texttt{arfgen} and \texttt{rmfgen}, respectively, 
for each ObsID. For MOS, effective area and response matrix files were computed for
each detector.  

\begin{table}
\caption{\textit{XMM-Newton} data: Total exposure times obtained for each ObsID after data reduction. 
} 
\label{table:XMM data}      
\centering          
\begin{tabular}{c  c  c  c  c }     
\hline   
\hline
ObsID & EPIC detector & $t_{exp}$  & RGS detector & $t_{exp}$ \\
      &               & [ks]                 &              & [ks]               \\
\hline  
     & MOS1 &  28  & RGS1 & 27 \\
    0148990101 & MOS2 &  29 & RGS2 & 27 \\
     & pn &  18  & - & - \\
    \hline
    \hline
     & tot EPIC & 75 & tot RGS & 54 \\
    \hline
    \hline
     & MOS1 &  113  & RGS1 & 118 \\
    0900280101 & MOS2 &  112 & RGS2 & 118 \\
     & pn &  98  & - & - \\
    \hline
    \hline
     & tot EPIC & 323 & tot RGS  & 236 \\
    \hline
    \hline
     & MOS1 & 97 & RGS1 & 116 \\
    0900280201 & MOS2 &  94 & RGS2 & 116 \\
     & pn &   50 & - & - \\
    \hline
    \hline
     & tot EPIC & 241 & tot RGS & 232 \\
    \hline
\end{tabular}
\end{table}

\subsection{XMM-Newton: RGS data}
\label{subsec:RGS data}

The RGS 1 and RGS 2 spectra were extracted using the centroid RA=23:51:39.42 and 
Dec = -26:05:03.02 and a width of 50 arcsec by adopting the mask 
\texttt{xpsfincl} = 90 in the \texttt{rgsproc} ($\sim$ $\pm$ 190 kpc at redshift z=0.23). 
In order to subtract the background, we tested both the standard background spectrum, which was extracted beyond the 98$\%$ of the RGS point spread function 
(PSF, \texttt{xpsfexcl} = 98 in the \texttt{rgsproc}), and the model background spectrum, 
which is a template background file based on blank field observations and the 
count rate measured in CCD 9. For extended sources, it is preferable to use the 
model background, but due to the small angular size of the CC region, 
we can safely use the observation background  extracted beyond 
a radius of $\pm 1.4$ arcmin from the BCG. 
From the analysis, we can see that the background spectra are indeed comparable 
and provide consistent results in the RGS 6–33 $\AA$ wavelength band.
The spectra were converted to the SPEX\footnote{www.sron.nl/spex} 
format through the SPEX task \texttt{trafo}. We focused on the first-order 
spectra because the second-order spectra have rather poor statistics. We also combined 
source and background spectra, and responses from all observations using the SAS task 
\texttt{rgscombine} to fit a single high-quality RGS spectrum and to speed up the fitting procedure.

\section{Spectral analysis of EPIC data}
\label{sec:spec EPIC}

\subsection{Analysis strategy}
\label{subsec:strategy}

We extracted the spectra from a region chosen to include the bulk of the
emission from the core region, which is estimated to be confined
within a radius of 40 kpc (about 11 arcsec at z $\sim$ 0.23), to consider the
effect \textit{XMM-Newton} PSF, whose half-power diameter is 7.5 arcsec, and, finally, 
to cover the same size of the extraction region used for RGS data, equal to 
$\pm$0.4 arcmin.  The combination of all these three requirements results in 
an optimal radius of 0.5 arcmin.

The background was sampled from a nearby region on the
same CCD that is free from the cluster emission, to be subtracted
from the source spectrum. We note that the total background
expected in the source region, computed by geometrically
rescaling the sampled background to the source area, only
amounts to 0.5$\%$ of the total observed emission, and therefore
statistical uncertainties in the background level are expected to have
a rather mild effect on the spectral fits.

\subsection{Isothermal collisional equilibrium model and cooling rate}
\label{subsec:spectral results EPIC}

Our analysis focuses on the 0.5-7 keV energy range of the
EPIC/MOS and EPIC/pn spectra, where the source counts are 
above the background and are not affected by XMM calibration issues. 
We performed the spectral analysis with SPEX version 3.08.00 \citep{1996Kaastra}. 
We scaled the elemental abundances to the proto-Solar abundances of \cite{2009LoddersPalme}, 
which are the default in SPEX, use C-statistics, and adopted 1$\sigma$ errors, 
unless otherwise stated. In the spectral modeling, we also included the central AGN 
whose spectrum has been measured with {\sl Chandra} high-resolution data. 

We describe the ICM emission with an isothermal plasma
model of collisional ionization equilibrium (\emph{cie}), where the free
parameters in the fits are the emission measure, $\rm Y=n_{e}n_{H}V$ (where $\rm n_{e}$ and $\rm n_{H}$ are the electron and Hydrogen densities, respectively, and V the volume of the source),
the temperature, T, and the iron abundance. The abundances of all the other elements 
are coupled to iron. Galactic absorption is set to 
$\rm N_{H}=1.54 \times 10^{20}$ $\rm cm^{−2}$ from \citet{2016HI4PI}.
We describe the central AGN with a power law (\emph{pow}) with $\Gamma = 1.8$, 
intrinsic absorption of $\rm N_{H}=1.37 \times 10^{23}$ $\rm cm^{−2}$, and $\rm log(L_{X})$[erg/s]=43.39,
as was obtained by our group from the analysis of the {\sl Chandra} data (see \citealp{2018Yang}). In this work, we do not account for AGN variability, as it can be considered negligible; however, we discuss its contribution to the fitting process in Appendix \ref{app:AGN variability}.

To place constraints on the amount of gas cooling below 3 keV down to 0.5 keV, we also added a CF component to the isothermal gas. 
The CF model in SPEX calculates the spectrum of a standard isobaric CF. 
The differential emission measure distribution for the isobaric CF model can be written as
\begin{equation}
    D(T) \equiv \frac{dY(T)}{dT} = \frac{5 \dot{M} k}{2 \mu m_{H} \Lambda(T)} 
,\end{equation}
where $\rm \dot{M}$ is the MDR, k is Boltzmann’s constant, 
$\rm \mu$ the mean molecular weight (0.618 for a plasma with 0.5 times solar abundances), 
$\rm m_{{H}}$ is the mass of a hydrogen atom, and $\rm \Lambda(T)$ is the cooling function \citep{2006Peterson}. 
The cooling function was calculated by SPEX for a grid of temperatures 
and for an average metallicity of $Z=0.5 Z_\odot$. The spectrum was evaluated by 
integrating the above differential emission measure distribution between 
a lower- and an upper-temperature boundary.

We used a model with three CF components plus the isothermal \emph{cie} all corrected by 
redshift and absorption. The aim is to understand how much gas is cooling between 
three temperature ranges (0.5–1.0, 1.0– 2.0, 2.0–3.0 keV). The only free parameter 
in each CF is the cooling rate, $\rm \dot{M}$, whilst the abundances are coupled 
to those of the \emph{cie} component, which are constrained by the iron line. We chose this approach based on the methodology presented in \cite{2016Molendi}. By using this multi-zone approach, we maximized the probability of detecting cooler gas and provided a more complete analysis of the CFs. Also, we chose this approach after testing that leaving the abundance parameters of the CF component free would result in unconstrained best-fit metallicity values. 
We are aware that our average metallicity is strongly driven by the iron abundance.

\subsection{Results}
\label{subsec:results_EPIC}
We present the results of our spectral analysis for each ObsID separately in Table 
\ref{tab:results EPIC fit}, where we show the best fit values of the MOS1+MOS2+pn joint fits. 
Our model provides a reasonably good fit of the MOS and pn spectra, as is shown by the Cstat values in the table. We show MOS1+MOS2+pn spectra with the best fit from the three different ObsID in Appendix \ref{app:EPIC spectra}. 

We also note that all the spectral best-fit values are
in agreement between independent ObsID. If we focus on the MDR in each
energy bin, we find substantial cooling between 2 and 3 keV, which is not surprising since
the bin is centered on a value close to 1/3 of the virial temperature ($\sim 7.5$ keV). 
Indeed, this value is similar to the one estimated from the total X-ray luminosity 
in the assumption of an isobaric CF model by \citet{2018McDonald}, equal to 
$\rm \dot M=580 \pm 180$ $\rm M_{\odot} yr^{-1}$.
Then, we only measure upper limits on the amount of cooling gas in the energy range 
1-2 keV. Interestingly, we nominally find 2 or 3 $\sigma$ detection
for a CF in the energy range 0.5-1 keV, at a level of $\sim 40$ $\rm M_\odot yr^{-1}$. 
Thus, our spectral analysis of CCD spectra suggests that the MDR below
2 keV is at maximum $\sim 3-5$\% of that corresponding to the isobaric CF model.

\begin{table*}
\caption{Results from EPIC/MOS1, EPIC/MOS2, and EPIC/pn joint fit for the three different ObsID.
}
\label{tab:results EPIC fit}
\centering
\begin{tabular}{l l l l l l l l}
\hline\hline
ObsID & Norm & Temperature & Abundance & $\rm \dot{M}$ (0.5-1keV) & $\rm \dot{M}$ (1-2keV) & $\rm \dot{M}$ (2-3keV) & Cstat/d.o.f\\
& [$10^{73}$ $\rm m^{-3}$] & [keV] & [$\rm Z_{\odot}$] & [$\rm M_{\odot}$ $\rm yr^{-1}$] & [$\rm M_{\odot}$ $\rm yr^{-1}$] & [$\rm M_{\odot}$ $\rm yr^{-1}$] & \\
\hline 
&&&&&&\\
0148990101 & $3.22 \pm 0.13$ & $7.71^{+0.45}_{-0.41}$ & $0.75^{+0.05}_{-0.04}$ & $46 \pm 20$ & $<21$ & $1170^{+97}_{-102}$ & 290/297 \\
&&&&&&\\
\hline
&&&&&&\\ 
0900280101 & $3.44 \pm 0.08$ & $7.63^{+0.23}_{-0.22}$ & $0.63 \pm 0.02$ & $45 \pm 11$ & $<18$ & $1129^{+58}_{-62}$ & 414/320\\
&&&&&&\\
\hline
&&&&&&\\
0900280201 & $3.43^{+0.14}_{-0.13}$ & $7.57^{+0.33}_{-0.30}$ & $0.61 \pm 0.03$ & $35 \pm 20$ & $<104$ & $1075^{+131}_{-133}$ & 408/313 \\
&&&&&&\\
\hline
&&&&&&\\
all stacked EPIC & 3.21 $\pm$ 0.05 & 7.64 $\pm$ 0.16 & 0.64 $\pm$ 0.02 & 46 $\pm$ 8 & <15 & $1127^{+39}_{-46}$ & 1101/709 \\
&&&&&&\\
\hline

\end{tabular}
\end{table*}


\section{Spectral analysis of RGS data}
\label{sec:spec RGS}

\subsection{Analysis strategy}
\label{subsec: Isothermal collisional equilibrium model}

Our analysis focuses on the 6.5−27  $\AA$ wavelength range of the first-order RGS spectra, 
where the source counts are above the observed background. This includes the 
K-shell emission lines of sulfur (S), silicon (Si), magnesium (Mg), neon (Ne), oxygen (O), and the 
L-shell lines of iron (Fe) and nickel (Ni). 

In this case, we describe the ICM and AGN emission with an isothermal plasma model of 
collisional ionization equilibrium and an absorbed power law model to model the AGN, 
as was already described in Sect.
\ref{subsec:spectral results EPIC}. The only difference with respect to the 
fit of the EPIC data is that here the abundances for O, Ne, Mg, and Si, whose dominant Lyman 
lines are clearly observed in the spectra, are considered free parameters, while 
the remaining elements are coupled to iron. Another major difference 
is that we minimize the fit with respect to the micro-turbulence 1D velocity 
broadening parameter. In addition, we also consider the redshift as a free parameter 
to take into account any eventual line shift. Other parameters, such as 
Galactic absorption and AGN intrinsic absorption are the same as in the EPIC spectra analysis
(see Sect. \ref{subsec:spectral results EPIC}). 

Eventually, to obtain better constraints on the amount of gas cooling below 0.5 keV, we fit the RGS spectrum by adding a CF model to the isothermal model, similarly 
to what we did for EPIC spectra in Sect. \ref{subsec:spectral results EPIC}. 
First, we considered three independent CF components in the temperature ranges 0.5–1.0, 1.0– 2.0, 
and 2.0–3.0 keV, plus the isothermal \emph{cie}.  
The only free parameter in each CF is the cooling rate, $\rm \dot{M}$, while 
the abundances are coupled to that of the \emph{cie} component.  As a second step, we considered a more complex 
model with a CF component with five temperature bins 
(0.1–0.5, 0.5–1.0, 1.0– 2.0, 2.0–3.0, and 3.0–7.0 keV). 
Finally, in order to place tighter constraints, we adopted a simplified model with only 
two temperature bins (0.1–2.0, 2.0–7.0 keV) for the CF component 
in addition to the isothermal gas model.

\begin{figure*}
   \centering
   \includegraphics[scale=0.60]{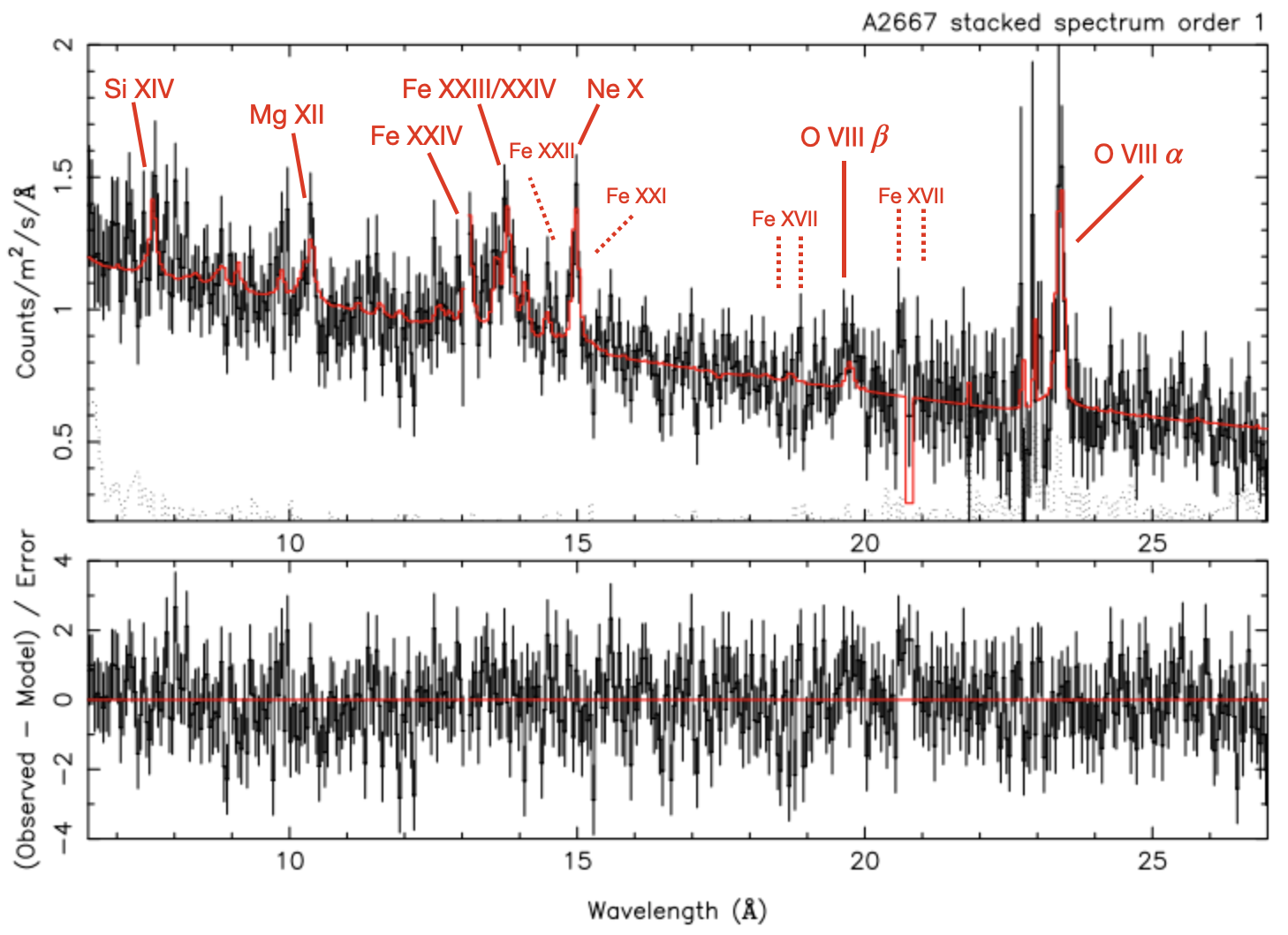}
   \caption{RGS spectrum of A2667 overlaid with an isothermal model of gas in collisional equilibrium (red). Emission lines commonly detected in RGS spectra of CC clusters are labeled at the observed wavelengths. Dotted labels refer to lines from gas below 1 keV. Residuals are shown in the bottom panel. The dotted line in the upper panel represents the background spectrum.
   }
    \label{fig:RGS stacked spectrum}
\end{figure*}

\begin{table*}
\caption{\textit{XMM-Newton}/RGS best-fit values for the isothermal \emph{cie} model. 
}
\label{tab:results RGS fit}
\centering
\begin{tabular}{l l l l l}
\hline\hline
$\rm n_{e}n_{H}V$ & Temperature & O/H & Ne/H & Mg/H \\
\hline
&&&&\\
6.70$\pm$0.08 & 4.27$^{+0.26}_{-0.24}$ & 0.47$ \pm 0.09$ & 0.51$\pm$ 0.09 & 0.37$\pm$0.13\\
&&&&\\
\hline
Si/H & Fe/H & vrms & z & Cstat/d.o.f \\
\hline
&&&&\\
0.31$\pm$0.12 & 0.49$^{+0.08}_{-0.07}$ & 710$\pm$170 & 0.2336 $\pm$ 0.0005 & 419/398\\
&&&&\\
\hline 
\end{tabular}
\tablefoot{Abundances are in proto-Solar units \citep{2009LoddersPalme}, emission measure $\rm n_{e}n_{H}V$ in $10^{73}$ $\rm m^{−3}$, temperature in keV and micro-turbulence 1D velocity in km $\rm s^{-1}$ (in this case representing the total width, i.e., spatial plus turbulent broadening).}
\end{table*}

\begin{figure}
   \centering
   \includegraphics[scale=0.55]{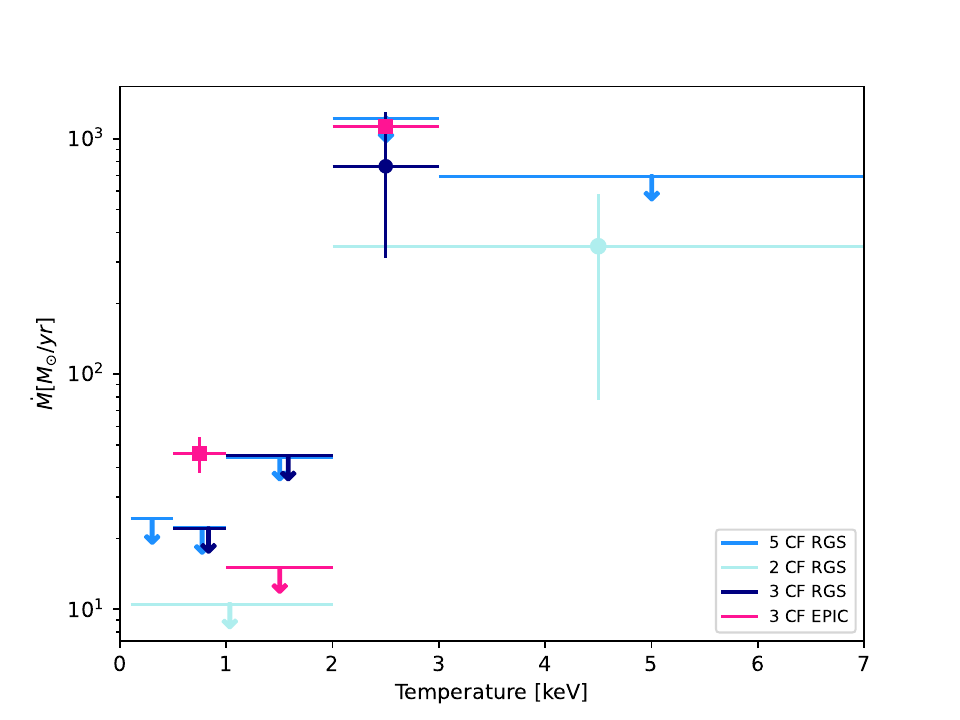}
   \caption{RGS and EPIC cooling rates measurements and upper limits. In light blue, dark blue, 
   and cyan, we show RGS values and upper limits on cooling rates obtained from a five-component, three-component and two-component CF model. In magenta, we show the EPIC measurements and upper limit of cooling rate obtained from the three-component CF model reported in Table \ref{tab:results EPIC fit} (see all stacked EPIC ObsID).}
    \label{fig:cooling rate}
\end{figure}

\subsection{Results}

The best-fit values for the simple isothermal plasma model of 
collisional ionization equilibrium are shown in Table \ref{tab:results RGS fit}. 
This model provides a reasonably good fit of the RGS spectra. 
Most strong lines are well described by the isothermal model, as is shown 
in Fig. \ref{fig:RGS stacked spectrum}. We detect several lines, particularly SiXIV, MgXII, FeXXIII/FeXXIV, NeX, and OVIII$\alpha$. We see that the spectrum lacks OVII, FeXXI-FeXXII, and FeXVII forbidden, recombination, and resonant emission lines.
We note that all the abundances of detected heavy elements are consistent with the common value Z=0.5$Z_{\odot}$, confirming our choice of using a single value for metallicity when fitting EPIC data. However, this 
value is at least 20\% lower than that obtained from EPIC data.  We also note that an isothermal \emph{cie} model is providing a good fit to the RGS spectrum despite a much lower temperature value with respect to the \emph{cie} component in EPIC data analysis. This suggests that the emission measure of the cold gas does not have a strong impact on RGS data. 

Not surprisingly, when we add a three-temperature bins (0.5–1.0, 1.0–2.0, 2.0–3.0 keV) CF model, we obtain 
only upper limits on the MDR, equal to 
$\rm \dot{M} (0.5-1 \rm keV)<22 M_{\odot} yr^{-1}$ and $\rm \dot{M} (1-2 \rm keV)<45$ $\rm M_{\odot} yr^{-1}$. 
As expected, the 2-3 keV temperature bin is characterized by a large cooling rate 
$\rm \dot{M} (2-3 \rm keV)=764^{+452}_{-542} M_{\odot} yr^{-1}$. These results are statistically 
in agreement with the ones found in the EPIC spectra. However, we do not reproduce the 2$\sigma$ 
detection of a CF in the 0.5-1 keV temperature bin.

Eventually, a more complex model with five temperature bins 
(0.1–0.5, 0.5–1.0, 1.0–2.0, 2.0–3.0, 3.0–7.0 keV) provide only upper limits on the 
amount of gas cooling within each temperature interval. We obtain $\rm \dot{M}(0.1-0.5 keV)<24$ $\rm M_{\odot} yr^{-1}$, $\rm \dot{M}(0.5-1.0 keV)<22$ $\rm M_{\odot} yr^{-1}$ and $\rm \dot{M}(1.0-2.0 keV)<44$ $\rm M_{\odot} yr^{-1}$.
Finally, when using a CF component with only two temperature bins (0.1–2.0, 2.0–7.0 keV), we obtain 
$\rm \dot{M}(2.0-7.0 keV)=350^{+232}_{-272}$ $\rm M_{\odot} yr^{-1}$, and  $\rm \dot{M}(0.1-2.0 keV)<11$ $\rm M_{\odot} yr^{-1}$. We report the MDR of the CF component with five, three and two temperature bins in Table \ref{tab:MDR RGS TOT}. 

To discuss our results, including the analysis of EPIC data, we show the measured MDRs or their 1 $\sigma$ upper limits in Fig. \ref{fig:cooling rate}.  First, we note that 
all our results are in agreement with each other, except the measurement of a mass deposition 
rate of $\rm \dot M = (46\pm8)$ $\rm M_{\odot}yr^{-1}$ in the temperature interval 0.5-1 keV obtained with 
the combined analysis of all the EPIC spectra. This point is, in fact,  about 3$\sigma$ above 
the upper limits found by the analysis of RGS spectra. Nevertheless, it is clear that the
maximum MDR allowed by an isobaric CF model is at least one order of magnitude
lower than that measured in the 2-3 keV temperature bin, in agreement with results obtained in 
similar studies \citep{2001Peterson,2010Sanders, 2016Molendi}. Given the overall agreement between the $\rm \dot M$ values, we shall perform a joint
EPIC and RGS fit in Sect. \ref{sec:spec_joint}.

\begin{table*}
\caption{\textit{XMM-Newton}/RGS best-fit MDRs for a CF component with five, three, and two temperature bins.}
\label{tab:MDR RGS TOT}
\centering
\begin{tabular}{l l l l l l}
\hline\hline
Model & $\dot{M}$(0.1-0.5keV) & $\dot{M}$(0.5-1.0keV) & $\dot{M}$(1.0-2.0keV) & $\dot{M}$(2.0-3.0keV) & $\dot{M}$(3.0-7.0keV) \\
\hline
&&&&&\\
5 CF & <24 & <22 & <44 & <1225 & <693 \\
&&&&&\\
\hline
Model & & $\dot{M}$(0.5-1.0keV) & $\dot{M}$(1.0-2.0keV) & $\dot{M}$(2.0-3.0keV) & \\
\hline
&&&&&\\
3 CF & & <22 & <45 & $765^{+452}_{-542}$ & \\
&&&&&\\
\hline
Model &  & $\dot{M}$(0.1-2.0keV)  & & & $\dot{M}$(2.0-7.0keV)\\
\hline
&&&&&\\
2 CF &  & <10  & & & $350^{+232}_{-272}$ \\
&&&&&\\
\hline 
\end{tabular}
\tablefoot{$\rm \dot{M}$ is in solar masses per year 
as the default in SPEX.}
\end{table*}

\subsection{Instrumental versus turbulent broadening}
\label{subsec: instrumental broadening}

The spectral fits yield a maximum line width corresponding to a line broadening 
$\rm v_{mic}$(1D)=710$\pm$170 km $\rm s^{-1}$. 
This measurement, however, does not account for the fact that the broadening is expected to be 
dominated by the spatial extent of the source. 
The impact of this effect can be subtracted using the surface brightness 
modeled on the high-resolution \emph{Chandra} images in the soft band [0.5-2 keV].

\begin{figure}
   \centering
   \includegraphics[scale=0.25]{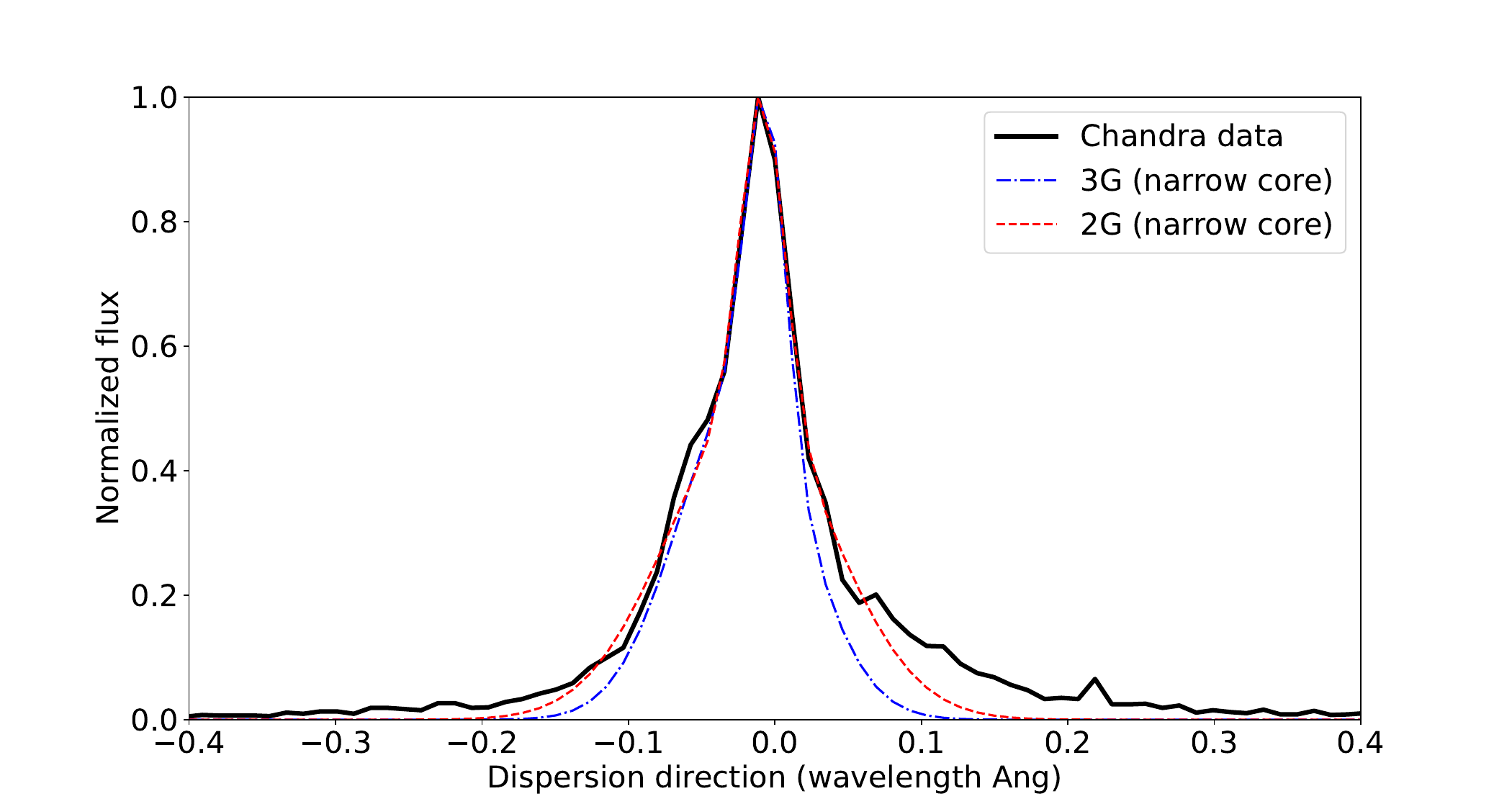}
   \caption{RGS line spatial broadening computed through the \emph{Chandra} surface brightness profile and Eq.\ref{eq: dispersion} (black line). The lines show the narrowest Gaussian referring only to the cluster core obtained by fitting two different models with either two (dotted red line) or three (dotted blue line) Gaussian lines.}
    \label{fig:cumulative}
\end{figure}

We produced \emph{Chandra} surface brightness profiles to model the RGS line spatial broadening with the standard dispersion equation:
\begin{equation}
    \Delta \lambda =\frac{0.138 \Delta \theta}{m} \AA
    \label{eq: dispersion}
,\end{equation}
where $\rm \Delta \lambda$ is the wavelength broadening, $\rm \Delta \theta$ is the source extent in arcseconds, and m is the spectral order (see the \textit{XMM-Newton} Users Handbook\footnote{http://xmm-tools.cosmos.esa.int/external/xmm$\_$user$\_$support/docu\-mentation/uhb/XMM$\_$UHB.pdf}). The surface brightness was extracted using a region of 50'' width and a length of 10' (which approximates an RGS extraction region;
see Sect.\,\ref{subsec:RGS data}), and using the \texttt{CIAO} task \texttt{dmstat}. The RGS instrumental line broadening measured through the \emph{Chandra} surface brightness profile of A2667 cluster is shown in Fig. \ref{fig:cumulative} (black line). 
It is well known that the spatial profiles extracted from the CCD images tend to overestimate the soft X-ray line broadening due to the contribution from the hotter continuum, which has a larger extent than the cooler gas producing such lines \citep{2015Pinto}.
We therefore followed the approach outlined in \cite{2018Bambic} whereby only the core of the spatial profile was adopted. This can be estimated by fitting the broadening profile with either two or three 
multiple Gaussian lines depending on the cluster distance. Typically, for clusters beyond redshift 0.2, two Gaussian lines are sufficient (i.e., one for the core and the other for the ambient gas). The single Gaussian line (black line) obtained with \emph{Chandra} data and the narrowest Gaussians (dotted green and red lines) of the multicomponent models fitting are also shown in Fig. \ref{fig:cumulative}, but renormalized for plotting purposes.

Then, we used the \emph{lpro} model in SPEX to take into account the instrumental broadening. The \emph{lpro} model receives as input the spatial broadening measured with the \emph{Chandra} surface brightness profile and convolves it with the baseline spectral model:
\begin{equation}
   Model=(hot \times (redshift \times cie)) \circledast lpro.
\end{equation}
The \emph{lpro} component now accounts for the spatial broadening.
As was mentioned before, we only accounted for the spatial extent of the 
CC in order to avoid over-subtraction of the spatial extent and under-estimate 
the upper limit on the turbulence. This was done by using the spatial broadening 
constrained by the narrower Gaussian that was used in the fit of the \emph{Chandra} 
spatial profile. After fitting the RGS spectrum including the \emph{lpro} component, the 
parameter $\rm v_{mic}$ is now related only to the turbulent velocity. The large uncertainty 
associated with the subtraction of the source profile hampers us from obtaining
a robust measurement. Indeed, we obtain a 90$\%$ upper limit of 561 km/s and 787 km/s 
for the velocity broadening in the case of 
two and three Gaussians, respectively.  We note that the velocity broadening upper limits 
are higher than those found in 
the Phoenix cluster at high redshift, and of those measured in other clusters 
in the local Universe, such as Abell 1835 \citep{2010Sanders}, observed with 
comparable signal to noise.  
This could be due to the fact that in A2667 there are some ongoing 
processes boosting turbulence more than in typical CC clusters. This point will be 
discussed in detail in Sects. \ref{subsec:cooling-heating equilibrium} and \ref{sec:minor mergers, sloshing, AGN feedback}. 

\section{EPIC and RGS joint fit}
\label{sec:spec_joint}

To better characterize the system and place tighter constraints on cooling rates, 
we performed a joint fit of all available \textit{XMM-Newton} spectra. We followed the strategy 
used for EPIC and RGS spectral fit separately in the previous sections.  
We described the ICM emission with an isothermal collisional equilibrium model, 
the AGN contribution with an absorbed power law, and we added a three-component CF model with 
temperature bins set to 0.5-1.0, 1.0-2.0, and 2.0-3.0 keV (see Sects. 
\ref{sec:spec EPIC} and \ref{sec:spec RGS} for further details). 

\begin{table*}
\caption{\textit{XMM-Newton}/RGS and EPIC joint best-fit isothermal \emph{cie} model with a CF component with three temperature bins. 
}
\label{tab:EPIC RGS joint fit}
\centering
\begin{tabular}{l l l l l l l}
\hline\hline
Instrument & $\rm n_{e}n_{H}V$ & Temperature & O/H & Ne/H & Mg/H & \\
\hline
&&&&&&\\
RGS + stacked MOS1&4.59$^{+0.16}_{-0.14}$ & 7.44$^{+0.32}_{-0.30}$ & 0.49$ \pm 0.08$ & 0.44$\pm$0.07 & 0.46$\pm$0.10 &\\
&&&&&&\\
RGS + stacked MOS2 & 4.97$\pm$0.15 & 7.12$^{+0.28}_{-0.26}$ & 0.53$ \pm 0.09$ & 0.40$\pm$0.07 & 0.39$\pm$0.10 &\\
&&&&&&\\
RGS + stacked pn & 4.40$\pm$0.13 & 7.82$^{+0.26}_{-0.24}$ & 0.47$ \pm 0.08$ & 0.40$\pm$0.07 & 0.44$\pm$0.09 &\\
&&&&&&\\
RGS + all EPIC & 4.45$\pm$0.10 & 7.80$^{+0.19}_{-0.18}$ & 0.48$ \pm 0.09$ & 0.37$\pm$0.06 & 0.45$\pm$0.07 &\\
&&&&&&\\
\hline\hline
Instrument & Si/H & Fe/H & $\rm \dot{M}$(0.5-1keV) & $\rm \dot{M}$(1-2keV) & $\rm \dot{M}$(2-3keV)& Cstat/d.o.f \\
\hline
&&&&&&\\
RGS + stacked MOS1 & 0.25$\pm$0.06 & 0.60$\pm$ 0.03 & 25$^{+19}_{-20}$ & $< 44$ & 1480$^{+106}_{-140}$ & 705/629\\
&&&&&&\\
RGS + stacked MOS2 & 0.30$\pm$0.06 & 0.60$\pm$ 0.02 & $< 8$ & $< 10$ & 1246$^{+108}_{-112}$ & 723/629\\
&&&&&&\\
RGS + stacked pn & 0.32$\pm$0.06 & 0.65$\pm$ 0.02 & 32$^{+20}_{-18}$ & $< 81$ 
& 1543$^{+124}_{-119}$ & 792/630 \\
&&&&&&\\
RGS + all EPIC & 0.30$\pm$0.04 & 0.65$\pm$ 0.01 & 20$^{+16}_{-18}$ & $< 58$ 
& 1556$^{+78}_{-102}$ & 1379/1094\\
&&&&&&\\
\hline 
\end{tabular}
\tablefoot{Abundances are in proto-Solar units \citep{2009LoddersPalme}, emission measure $\rm  n_{e}n_{H}V$ in $10^{73}$ $\rm m^{−3}$, temperature in keV and $\rm \dot{M}$ in solar masses per year as default in SPEX. }
\end{table*}

The best-fit values of the spectral parameters from the joint fit are shown in Table 
\ref{tab:EPIC RGS joint fit}, while we report the best-fit model and data plots in Apprendix \ref{app:EPIC and RGS joint fit spectra}.
We performed the joint fit in four steps: first, we combined RGS with the stacked spectrum of the MOS1 detectors, then MOS2, pn, and finally all the EPIC detectors together.  
We note that all the combinations are in agreement with each other, so we can rely on the 
fit obtained by stacking all the independent spectra extracted from our data as the result with the
largest amount of information. 
We conclude that we find positive values within 1$\sigma$ sigma confidence level for the MDR in the 0.5-1 keV temperature bin and only upper limits in the 1-2 keV temperature bin. These values and upper limits are in agreement with the analysis of the RGS spectrum within 1$\sigma$, and more in general with the values shown in Fig. \ref{fig:cooling rate}.


Finally, we note that the isobaric cooling may not be the only physical mechanism producing a
MDR. Therefore, after having explored CF models in a limited temperature bin, 
we account also for the presence of nonisobaric modes by the \emph{wdem} \citep{2004Kaastra}
model which describes the emission measure distribution of the ICM with a generic power law. 
The free parameters in this model are the emission measure value, maximum temperature, 
lower temperature cutoff, and iron abundance. The values found for the emission measure 
($\rm n_{e}n_{H}V=(6.60^{+0.05}_{-0.04})\times10^{73}$ $\rm m^{-3}$), 
maximum temperature ($\rm T_{max}=6.30\pm 0.03$ keV) and iron abundance 
($\rm Fe/H=0.62\pm 0.01$) are in agreement with the values found with \emph{cie} model, 
but with a much higher C-stat value (C-stat=1927/1100). Moreover, 
we are not able to constrain the value of the low-temperature cutoff 
($\rm T_{cut}<0.13$ keV). We conclude that the gas 
distribution in this cluster does not follow a power law, as has also been found 
by a similar analysis in the literature (see \citealp{2013Frank}).

\section{Discussion}
\label{sec: discussion}

\subsection{Cooling rates}
\label{subsec:cooling rates}

In this work, we study the galaxy cluster A2667 in order to understand how much gas is cooling 
below 2 keV. Considering the fitting procedure of the RGS and EPIC data, we can only put 
an upper limit on the values of cooling rate in the temperature ranges 0.5-1 keV and 1-2 keV. 
In particular,  in the range 0.5-1 keV, we have a 1$\sigma$ upper limit of $\sim$ 36 $\rm M_{\odot} yr^{-1}$, 
while in the range of 1-2 keV we have an upper limit of $\sim$ 60 $\rm M_{\odot} yr^{-1}$. 
On the one hand, in the RGS spectrum we only find evidence of the OVIII line directly associated
to $\sim 1$  keV ICM. On the
other hand, we do not significantly detect either FeXXI-FeXXII and FeXVII forbidden, 
intercombination, and resonant emission lines, or OVII, which are the best tracers 
for $\lesssim 1$ keV gas (see e.g., \citealp{2010Sanders}).  
Probably, the OVII lines are hidden within the noise and the high instrumental background 
around 35$\AA$. However, the absence of FeXXI-FeXXII should clearly indicate that the amount of gas 
cooling below 1 keV is minimal. 
Also, since the difference measured in the cooling rates  between 1-2 keV and 2-3 keV 
is larger than 1000 $\rm M_{\odot} yr^{-1}$, it looks like the gas is not able 
to cool below 2 keV because of reheating by some feedback mechanism.
As is supported by high-resolution hydrodynamical simulations, the sudden drop of
the emission measure below 2 keV is likely due to a combination of the direct AGN heating 
and turbulent mixing, which induce nonisobaric modes into the (quenched) 
CF (\citealt{2015Gaspari}).

These results, combined with the presence of an AGN with a low nuclear power 
($\sim$ 3 $\times 10^{43}$ erg/s, \citealp{2018Yang}), could imply that we are either 
in the presence of a minimal MDR in the aftermath of a cooling event 
powering off, or the beginning of a cooling episode that has just awakened the AGN.  In this second case, 
the BCG may evolve within a few tens of millions of years into a Phoenix-like phase, 
with a rise in the value of the MDR promptly counterbalanced by strong feedback. 
A relevant quantity in this respect would be an estimate of average AGN feedback power in the last tens of Myr, for example from ICM cavities. Moreover, it would be important to assess the dynamical 
state of the clusters to evaluate the presence of induced sloshing in the ICM or to identify
a recent minor merger onto the BCG. We shall discuss some of these aspects in the next subsections.

\subsection{Cooling-heating equilibrium}
\label{subsec:cooling-heating equilibrium}

In our attempt to place constraints on turbulence and estimate whether the energy stored 
in turbulent motions is high enough to balance radiative cooling, we obtained a
total broadening of the emission lines in the RGS spectrum of 710$\pm$170 km $\rm s^{-1}$. 
As is described in Sect. \ref{subsec: instrumental broadening}, after modeling the spatial broadening 
due to the spatial extent of the source with two Gaussian curves, 
we obtain 90\% upper limits on the turbulent broadening, 
equal to 561 km $\rm s^{-1}$, or even higher for more accurate modelization of the source profile.
This limit is well above the turbulence level required to propagate the heat 
throughout the Phoenix cluster at z$\sim$0.6 from the central AGN up to 250 kpc \citep{2018Pinto}. 
This suggests that in A2667 the ICM may be characterized by a high level of turbulence, 
consistent with a post-feedback scenario. 

To achieve a rough estimate of the turbulent velocity in the ICM of A2667, 
we can adopt the simplified, model-dependent approach proposed by \cite{2013Sanders}. 
This consists of computing the intrinsic broadening as the difference between the square
of the total width obtained from the RGS spectral fit (710 km/s) and the instrumental 
broadening estimated simply by the full width half maximum of the Gaussian 
describing the core spatial extent. The broadening in units of velocity 
is obtained from the relation
\begin{equation}
    \Delta v= \frac{\Delta \lambda}{\lambda} c = \frac{\sigma c}{\lambda}\, ,
    \label{eq:velocity}
\end{equation}
where $\sigma=0.04$ is the Gaussian width obtained with the Python routine {\tt leastsq}, $\lambda$ is the RGS band center at $\sim$19 $\AA$ (position of O VIII emission line peak), and \emph{c} is the speed of light. We obtain an intrinsic broadening of $320 \pm 200$ km $\rm s^{-1}$. 
However, this result is an optimistic estimate of the error. If we consider all the uncertainties in our approach, this result can be considered as a 1$\sigma$ upper limit. We have to rely on future bolometer measurements (e.g., XRISM; see Sect. \ref{sec:future perspectives}) to provide greater precision and allow us to confirm or refine this estimate.

As has been probed by high-resolution hydrodynamical simulations, such levels of turbulent velocities 
(and related enstrophy) are enough to enhance local thermal instability and trigger a 
condensation rain of warm filaments or clouds, which is expected to boost the SMBH accretion 
rates via CCA during the next self-regulation cycle 
(\citealt{2020Gaspari}). Indeed, the SMBH mass of A2667 is  $\rm 2\times10^{10} M_{\odot}$ 
\citep{2019Phipps}, which can be achieved only through recurrent flickering cycles of CCA feeding, 
unlike in hot-mode (Bondi/ADAF) scenarios (see the Discussion section in \citealt{2019Gaspari}).
Quantitatively, we computed a key CCA diagnostics; that is,~the condensation ratio, 
$C\equiv t_{\rm cool}/t_{\rm eddy}$, within the putative bubble region ($L\sim25$ kpc), 
which is also comparable to typical condensation core sizes (\citealt{2023Wang}). 
We find a $C$ ratio of 0.4, which is consistent with an ensuing or resuming CCA rain, 
as it crosses the $C\sim 1$ threshold (\citealt{2018Gaspari}).

\subsection{Comparison with Abell 1835}

To have a better grasp on the level of turbulence in A2667, we confronted the intrinsic broadening value obtained 
for A2667 with the one obtained from Abell 1835 (hereafter A1835) \emph{Chandra} surface brightness profile. 
We used this cluster because is placed at a redshift close to that of A2667 -- $z=0.2523$ --
and shows a CC at the center. 
In addition, considering the total luminosity of both
clusters, their temperature and the \textit{XMM-Newton} exposure time, the X-ray signal from A1835 is about 4.5 times larger than A2667. 
A1835 is the most luminous cluster in the \emph{ROSAT} Bright Cluster Sample \citep{1998Ebeling}, with an inferred isobaric CF 
cooling rate of $\rm \sim 1000$ $\rm M_{\odot} yr^{-1}$ \citep{1996Allen}. The SFR in the central galaxy 
derived from optical spectroscopic is 40–70 $\rm M_{\odot} yr^{-1}$ \citep{1999Crawford}. 
The infrared luminosity of the object is $\sim 7 \times 10^{11}$ $\rm L_{\odot}$ \citep{2006Egami}, 
suggesting an even higher SFR up to 
$\sim 125$ $\rm M_{\odot} yr^{-1}$. In addition a total mass of $\sim 5\times 10^{10}$ $\rm M_{\odot}$ 
of molecular gas has been measured within 10 kpc from the BCG center \citep{2014McNamara}.
Despite the evidence for star formation, cool gas and the high X-ray luminosity 
($L_{X[0.1-2.4 keV]}\sim$ 3.9 $\times 10^{45}$ erg $\rm s^{-1}$; \citealp{1998Ebeling}), 
little X-ray emitting gas was seen in this cluster below 1–2 keV 
when it was observed using the \textit{XMM-Newton} RGS instruments \citep{2001Peterson}. More quantitatively, 
fewer than 
200 $\rm M_{\odot} yr^{-1}$ cools below 2.7 keV (90$\%$ confidence) and far 
below 
100 $\rm M_{\odot} yr^{-1}$ at lower temperatures \citep{2019Liu}. Therefore, it has been 
proposed that AGN feedback, estimated to have an average of $1.4 \times 10^{45}$ erg/s in the last few tens of millions of years, 
prevents most of the X-ray gas from cooling \citep{2006McNamara}. 

Using XMM-RGS data, \cite{2010Sanders} calculate the line-of-sight, nonthermal, total broadening of A1835, 
finding a 90$\%$ upper limit value of 222 km/s. This value is lower than our measurement of the 
total broadening in A2667 equal to $710\pm 160$ km $\rm s^{-1}$ (see Sect. 
\ref{subsec: Isothermal collisional equilibrium model}). We attempt to estimate
the turbulent velocity component in A1835 after computing the source spatial broadening in the 
simplified method described in Sect. \ref{subsec:cooling-heating equilibrium}. 

\begin{figure}
   \centering
   \includegraphics[scale=0.25]{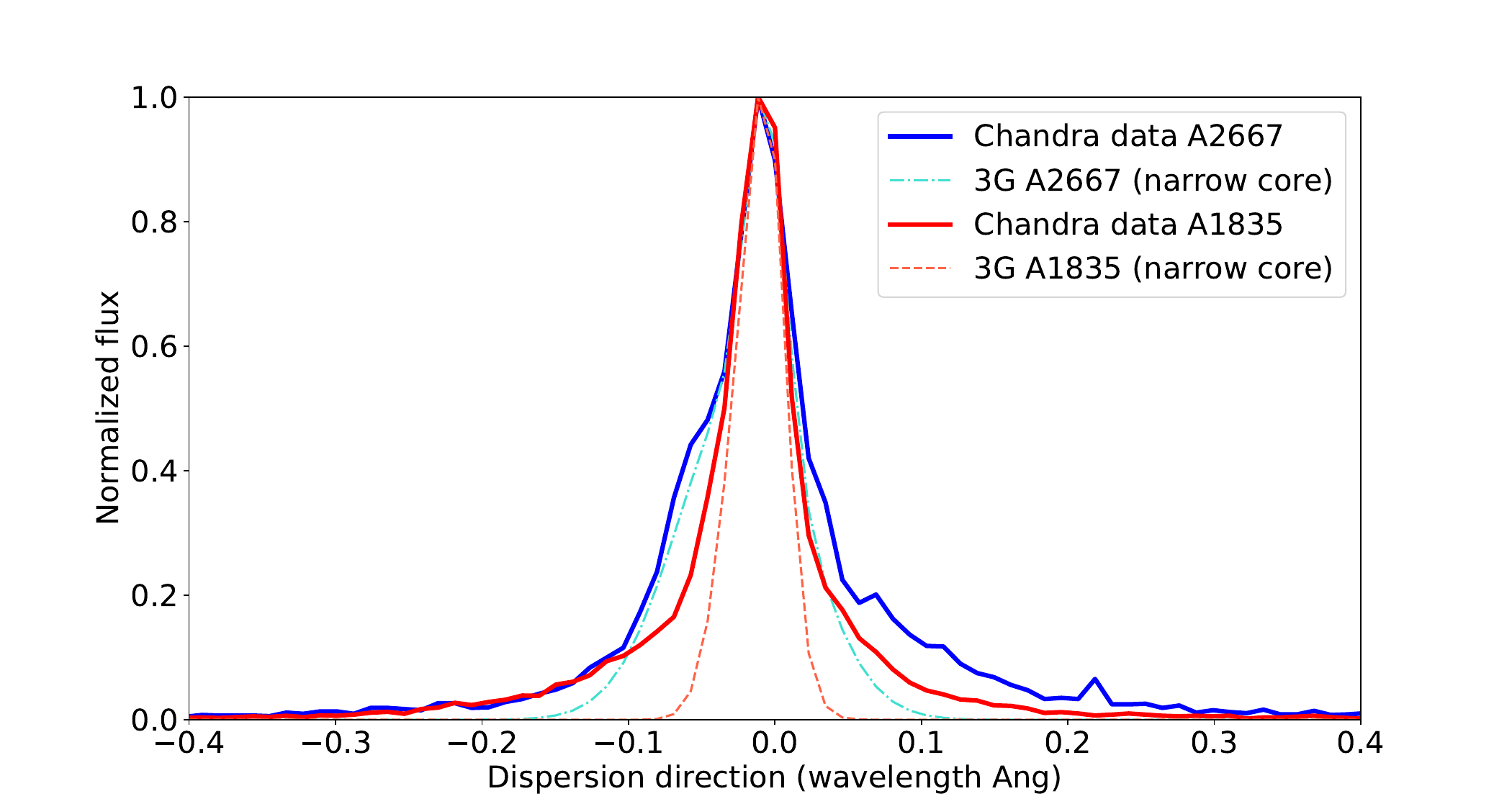}
   \caption{Line spatial broadening computed through the \emph{Chandra} surface brightness profile and Eq. \ref{eq: dispersion} for A2667 (blue line) and Abell 1835 (red line). The dash-dotted turquoise line and the dashed orange line show the narrowest Gaussian obtained by fitting three Gaussian lines for A2667 and Abell 1835, respectively.  }
    \label{fig:confronto A1835 A2667}
\end{figure}

We used 30 ks of archival ACIS data from the \emph{Chandra} archive to derive the surface brightness profile of A1835. 
We reduced the data following the procedure described in Sect. \ref{sec:chandra}.  We obtained the line spatial broadening 
fitting the data with three Gaussian lines (see red and orange lines in Fig. \ref{fig:confronto A1835 A2667}), 
as described in Sect. \ref{subsec: instrumental broadening}.  Then, we applied the relation \ref{eq:velocity}, 
where $\sigma=0.01$ was obtained by the Python routine {\tt leastsq}. We obtained an upper limit on the 
velocity broadening of 170 km/s (90\% confidence level) that can be ascribed to turbulent motions. 

The striking difference between the line broadening 
of A2667 with respect to A1835 can be seen in Fig. \ref{fig:confronto A1835 A2667}.
We note that the spatial broadening in A1835 is $\sim$ 50$\%$ lower than the spatial broadening in A2667 
and this reflects on the turbulent velocity line broadening which is at least double in A2667 
with respect to A1835.  It has been argued that the central AGN in A1835 
is not inducing large amounts of turbulence in the ICM in the cluster core \citep{2010Sanders}.
Given our results, we suggest that in A2667 turbulent velocity may be boosted by physical processes such as AGN mechanical feedback, sloshing, and/or minor mergers, or, more likely, a combination of them. We tentatively quantify this effect to be a factor of 2 with respect to A1835
and other similar CC clusters (see clusters in \citealp{2013Sanders} sample). 
To summarize, it will be extremely interesting to obtain tighter constraints on the turbulence 
in A2667 with X-ray bolometers, as is discussed in Sect. \ref{sec:future perspectives}.

\subsection{Possible role of minor mergers, sloshing, and AGN feedback in heating processes}
\label{sec:minor mergers, sloshing, AGN feedback}

In previous sections, we mentioned minor mergers, sloshing and AGN feedback as possible mechanisms 
for boosting turbulent velocity and reheating the ICM preventing or strongly reducing cooling in the
gas below 2 keV in A2667. Considering the resolution of current data, we cannot rule out one mechanism or 
identify the dominance of another. Here, we review different hints on the role of these mechanisms based on
the available literature. Indeed, according to previous studies, especially \cite{2019Giacintucci} 
and \cite{2019Iani}, there is convincing evidence of the presence of all these mechanisms. 

\begin{figure}
   \centering
   \includegraphics[scale=0.35]{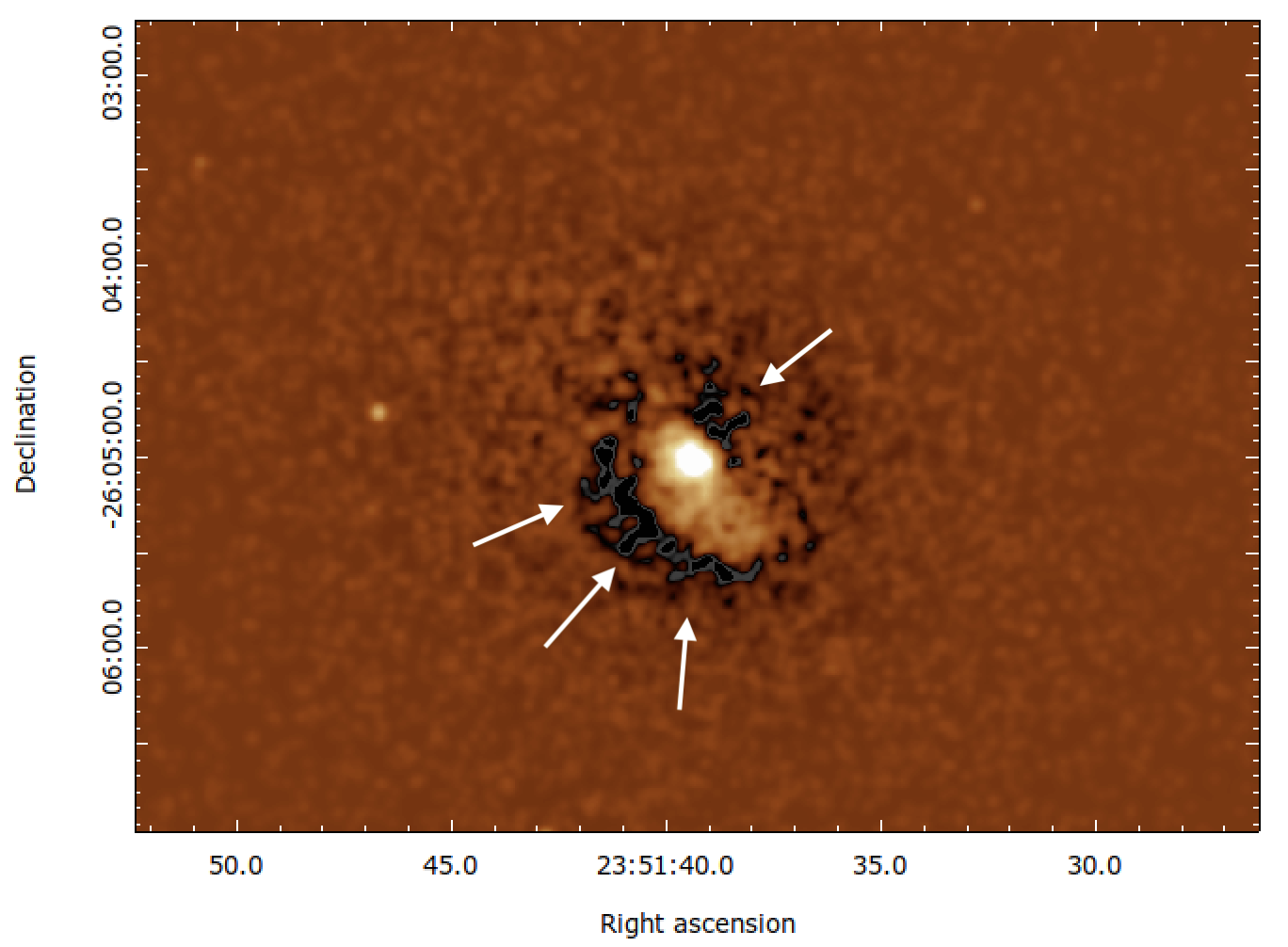}
   \caption{\emph{Chandra} 0.3-10 keV unsharp mask image of A2667. The white arrows are adapted from \cite{2019Giacintucci} and represent the two surface brightness edges found, which may be cold fronts or bubble-like features.}
    \label{fig:Chandra unsharp A2667}
\end{figure}

\cite{2019Iani} find the presence of a complex system of substructures in the optical 
surface brightness extending all around the BCG at the center of A2667 using HST and MUSE data, 
characterized by a strong blue optical color suggesting a young stellar population. They propose two scenarios 
to explain the presence of filaments and clumps, one associated with the AGN feedback and the other associated 
with the presence of a companion galaxy falling onto the BCG. Particularly, for the second scenario the clumps 
appear to lie on the BCG galactic plane, as it would happen with an accreting satellite galaxy that loses 
its angular momentum and starts inspiralling toward the BCG center. At the opposite extremity of the galaxy 
with respect to the clumps in the HST images, diffuse blue emission is present, likely due to the gas left 
behind by the satellite galaxy from a previous orbit. The presence of a satellite galaxy could favor the presence 
of minor mergers which influence the heating and cooling of the gas around the BCG and in the cluster. 
Instead, they tend to exclude the first scenario on the AGN feedback because even though the clumps are 
observed on a physical scale of $\sim$ 10–20 kpc from the galaxy nucleus, \emph{Chandra} X-ray data and 
JVLA-ALMA radio data do not show a clear presence of jets or inflated bubbles on a large scale. However, 
if we improve the contrast of \emph{Chandra} image using the \texttt{unsharp mask} procedure in \texttt{CIAO}, 
we find clear asymmetric features at a distance of roughly 90 kpc and 50 kpc 
from the center of the BCG, as is shown in Fig. 
\ref{fig:Chandra unsharp A2667}. 
The shape of the features is reminiscent of a cold front, however, the surface brightness 
depression regions are quite extended, and we cannot exclude the presence of cavities. 
Given the very shallow archival exposure, and the high flux of A2667, a moderately
deeper \emph{Chandra} observation would significantly improve the characterization of the 
surface brightness edges or cavities detected in current data. This would bring another piece of information that is relevant for identifying the main source of heating in the core of A2667. 

Focusing on the radio band, \cite{2019Giacintucci} find the presence of a 
minihalo using GMRT-VLA radio data which covers an area of $\sim$ 70 kpc in radius. 
They also discuss the two surface brightness features in the X-ray image 
(see white arrows in Fig. \ref{fig:Chandra unsharp A2667}) and interpret them as cold fronts. 
Indeed, the radio minihalo seems to be contained within the sloshing region 
defined by the X-ray fronts. These results stress the presence of sloshing in the ICM 
of A2667. To conclude, the combination of 
high-resolution X-ray spectra and high-resolution X-ray images is required to 
comprehend the dynamical and thermodynamical state of the ICM in the core of A2667.

\section{Future perspectives}
\label{sec:future perspectives}

Current X-ray instruments on board \textit{XMM-Newton} and
\emph{Chandra} only allow us to obtain approximate turbulent velocity measurements 
and also have limitations due to source extent and associated broadening. 
The most accurate measurement of line broadening and, therefore, constraint on turbulence in the ICM, 
was obtained by the Hitomi satellite \citep{2010Takahashi} during its deep (230ks) 
observation of the Perseus cluster of galaxies \citep{2016Hitomi}, but this satellite 
was lost immediately afterward in 2016. However, a new X-Ray Imaging and Spectroscopy 
Mission (XRISM; \citealp{2019GuainazziTashiro}) was launched in 2023. 

We therefore performed a simulation of A2667 as it will be seen by XRISM 
using a gate-valve-closed (GVC) Resolve response matrix with a 5 eV spectral 
resolution using SPEX command \texttt{simulate}. We considered 
the XMM/EPIC+XMM/RGS best fit model (\emph{cie} model; see Sects. 
\ref{sec:spec EPIC} and \ref{sec:spec RGS}) and assumed an exposure time of 250 ks with GVC, 
similar to that of the Perseus cluster. 

\begin{figure}
   \centering
   \includegraphics[scale=0.3]{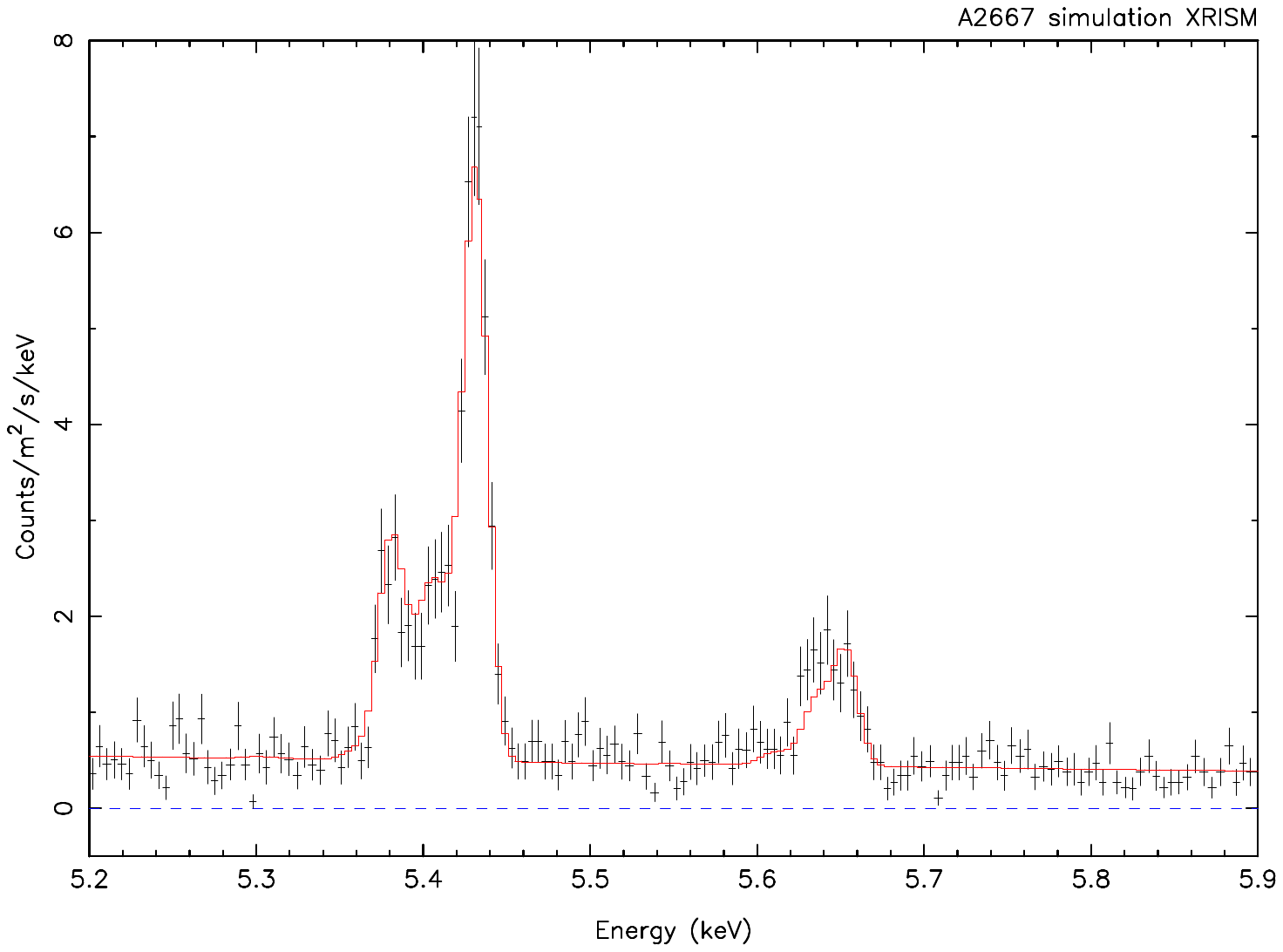}
   \caption{Simulation of an A2667 observation with XRISM / Resolve accounting for the GVC. The black dots are data points and the red line is the best-fit isothermal model. The plot is zoomed on the region of the FeXXV-XXVI lines in the observer frame.
   }
    \label{fig:simulazione XRISM}
\end{figure}

We assume a line broadening due to turbulent velocity of 320 km $\rm s^{-1}$ 
as in our simple estimate  (see Sect. \ref{subsec:cooling-heating equilibrium}). 
The simulated spectrum is shown in Fig. \ref{fig:simulazione XRISM}. 
We can see that we are able to resolve the Fe XXV multiplet emission lines
in the energy range of 5-6 keV. We can place tight constraints on turbulent
velocity corresponding to a statistical uncertainty of just $\sim$ 30 km $\rm s^{-1}$ 
for a 250 ks XRISM observation, obtaining much stronger
constraints compared to those in this work.

Another instrument that will bring a drastic improvement is the X-ray Integral Field Unit 
(X-IFU) on board the Advanced Telescope for High Energy Astrophysics
(ATHENA; \citealp{2013Nandra}) that has been rescoped to the novel 
mission design NewAthena and is expected to be operating in the late 2030s. 
This satellite will provide a spatial resolution
down to around 9 arcsec (half energy width), a spectral resolution of $\sim 2.5$ 
eV, and an effective area of more than an order of magnitude
larger than any previous grating or microcalorimeter, 
implying a significant improvement in our capabilities to investigate the
X-ray spectra of galaxy clusters. We performed a NewAthena simulation of A2667 
using a gate-valve-open (GVO) X-IFU response matrix (as of July 2024). 
We considered the same template model used for XRISM simulation with the 
addition of a CF model with three temperature bins, but with an exposure time of 300 ks, higher than the 250 ks used for the XRISM simulation.

\begin{figure}
   \centering
   \includegraphics[scale=0.3]{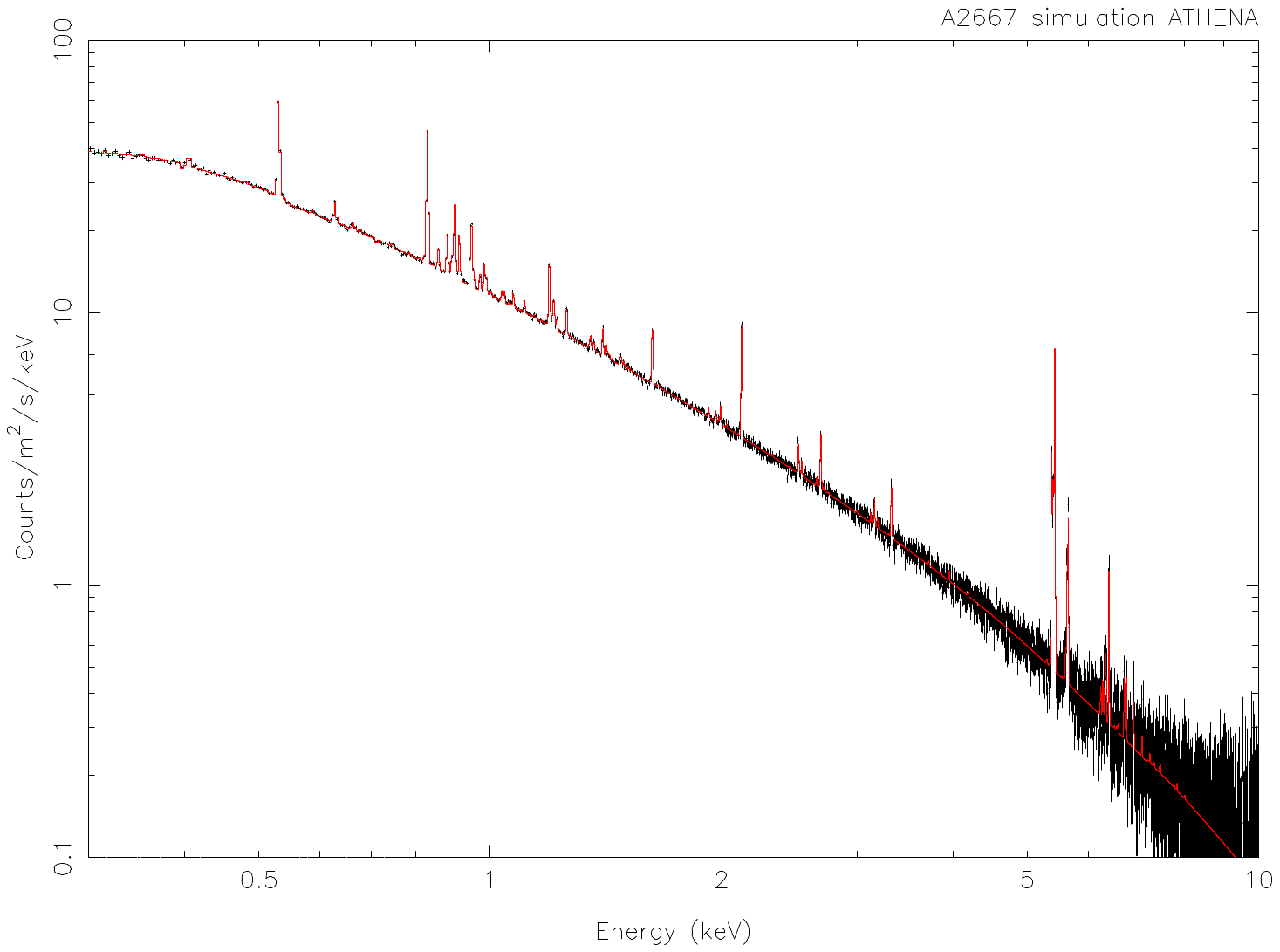}
   \caption{Simulation of an A2667 observation with ATHENA / X-IFU. The black dots are data points and the red line is the best-fit isothermal model + isobaric CF model in three temperature bins.}
    \label{fig:simulazione ATHENA}
\end{figure}

The X-IFU 300 ks observation of A2667 will provide a spectacular spectrum of high quality (see Fig. \ref{fig:simulazione ATHENA}) and will allow us to measure velocity broadening with
statistical uncertainties of $\sim$ 6 km $\rm s^{-1}$, much tighter constraints on turbulence compared to the XRISM observations. This also happens for cooling rates measurements, for which we obtain statistical uncertainties of $\sim 10-20$\% (0.5-1.0 keV) on an MDR of 
10 $\rm M_{\odot} yr^{-1}$. For the temperature bin 1.0-2.0 keV, on the other hand, it is still difficult to get this level of uncertainty. Indeed, to obtain statistical uncertainties of $\sim 10-20$\% on the MDR in the range 1.0-2.0 keV we need observations of 3 Ms. 

With NewAthena, we are also able to put constraints on the temperature truncation 
value of the cooling gas. In a simulation with an input isobaric CF with a minimum 
temperature of 0.7 keV, and $\rm \dot M = 100$ $\rm M_{\odot} yr^{-1}$, we are able to recover the truncation temperature with a 1$\sigma$ error of 4$\%$, and the MDR 
with a 1$\sigma$ error of 8$\%$.
We conclude that with XRISM and NewAthena we shall be able to investigate the cooling and heating mechanisms in A2667, and finally
constrain the different scenarios outlined in this work.

\section{Conclusions}
\label{sec:Conclusions}
In this paper, we have studied the thermal X-ray emission from the ICM in the CC cluster 
A2667 at redshift $z=0.233$, focusing on the core region included within a radius of $\sim$ 115 kpc. 
We performed a detailed analysis of the deep \textit{XMM-Newton} observations of A2667, 
along with archival \emph{Chandra} data. 

From the X-ray spectral analysis of XMM/EPIC and RGS data, we conclude that we are 
able to put only upper limits on the MDR in the ICM below 2 keV. In particular, we 
find $\rm \dot{M} < 36$ $\rm M_{\odot} yr^{-1}$ (0.5-1 keV) and $\rm \dot{M} < 58$ $\rm M_{\odot} yr^{-1}$ (1-2 keV). We measure a cooling rate in the temperature interval 2-3 keV equal to 
$\rm \dot{M}=1556^{+78}_{−102}$ $\rm M_{\odot} yr^{−1}$. 
These results envisage a scenario in which the maximum amount of gas
cooling below 2 keV is below $\sim$ 4-5$\%$ of gas cooling in the range 2-3 keV. 
Thus, we can say that A2667 is not different from other low-redshift CC clusters.

Interestingly, from XMM/RGS spectral fitting, using a \emph{cie} model, we find a 
maximum line broadening corresponding to $710 \pm 170$ km $\rm s^{−1}$,
which includes both spatial and turbulent broadening, along with instrumental broadening. 
We make a simple estimate of the intrinsic, turbulent velocity of $\sim 320\pm 200$ 
km $\rm s^{-1}$, a value that is at least twice the one estimated in 
the local cluster A1835. These limits are well above the turbulence required 
to propagate the heat throughout the Phoenix cluster at $z\sim 0.6$ from the central AGN 
up to 250 kpc. This result could be considered as a hint of the fact that there are 
some processes ongoing in A2667 that boost turbulence and quench cooling, such 
as sloshing, minor mergers, and/or AGN feedback. 
At the same time, it is important to note that turbulence stimulates local thermal 
instability, and hence vitally promotes the next generation of CCA rain, as is shown by 
the computed low $C$ ratio of 0.4 in the core of A2667.  Imaging analysis with high-resolution 
{\sl Chandra} data provide clear surface brightness depression that can be interpreted 
as the signatures of a cold front or the presence of ICM cavities.

Current XRISM and future NewAthena missions will
improve the accuracy in measuring crucial parameters in X-ray
spectra of clusters of galaxies. Particularly, it is already possible to 
bring the uncertainty on the turbulent velocity in A2667 down 
to $\sim 10$\% ($\sim$ 30 km $\rm s^{−1}$ in our case) with a 250 ks XRISM observation.
With NewAthena it will be possible to further reduce this uncertainty to 
$\sim 3$\%. Unfortunately, a robust assessment of the actual MDR
at the level of $\sim 10$ $\rm M_{\odot} yr^{-1}$, as we expect in A2667, will be possible 
only with an X-ray bolometer with sensitivity below 2 keV, such as the X-IFU on board NewAthena.  This implies that the characterization and quantification of the 
CFs hosted by CC clusters directly measured from X-ray is at present
severely limited by current facilities.

\begin{acknowledgements}
We thank the anonymous referee for the detailed comments and positive criticism that helped us improve the quality of this manuscript.
M.L. and P.T. acknowledge the program INAF 
Fundamental Astrophysics 2023 "The origin of cool cores and the evolution of BCGs in galaxy clusters". M.L. further acknowledges the support from the ESO Scientific Visitor Programme.
M.L. also acknowledges support from the PRIN MIUR 2020 GRAAL 2020SKSTHZ. C.P. is supported by PRIN MUR 2022 SEAWIND 2022Y2T94C funded by European Union - Next Generation EU.
M.G. acknowledges support from the ERC Consolidator Grant \textit{BlackHoleWeather} (101086804).

\end{acknowledgements}
      

\bibliographystyle{aa}
\bibliography{References_Abell2667.bib}

\begin{thebibliography}{95}
\expandafter\ifx\csname natexlab\endcsname\relax\def\natexlab#1{#1}\fi

\bibitem[{{Allen} {et~al.}(1996){Allen}, {Fabian}, {Edge}, {Bautz}, {Furuzawa}, \& {Tawara}}]{1996Allen}
{Allen}, S.~W., {Fabian}, A.~C., {Edge}, A.~C., {et~al.} 1996, \mnras, 283, 263

\bibitem[{{Bambic} {et~al.}(2018){Bambic}, {Pinto}, {Fabian}, {Sanders}, \& {Reynolds}}]{2018Bambic}
{Bambic}, C.~J., {Pinto}, C., {Fabian}, A.~C., {Sanders}, J., \& {Reynolds}, C.~S. 2018, \mnras, 478, L44

\bibitem[{{Cavagnolo} {et~al.}(2009){Cavagnolo}, {Donahue}, {Voit}, \& {Sun}}]{2009Cavagnolo}
{Cavagnolo}, K.~W., {Donahue}, M., {Voit}, G.~M., \& {Sun}, M. 2009, \apjs, 182, 12

\bibitem[{{Cavaliere} \& {Fusco-Femiano}(1978)}]{1978Cavaliere}
{Cavaliere}, A. \& {Fusco-Femiano}, R. 1978, \aap, 70, 677

\bibitem[{{Covone} {et~al.}(2006{\natexlab{a}}){Covone}, {Adami}, {Durret}, {Kneib}, {Lima Neto}, \& {Slezak}}]{2006bCovone}
{Covone}, G., {Adami}, C., {Durret}, F., {et~al.} 2006{\natexlab{a}}, \aap, 460, 381

\bibitem[{{Covone} {et~al.}(2006{\natexlab{b}}){Covone}, {Kneib}, {Soucail}, {Richard}, {Jullo}, \& {Ebeling}}]{2006aCovone}
{Covone}, G., {Kneib}, J.~P., {Soucail}, G., {et~al.} 2006{\natexlab{b}}, \aap, 456, 409

\bibitem[{{Crawford} {et~al.}(1999){Crawford}, {Allen}, {Ebeling}, {Edge}, \& {Fabian}}]{1999Crawford}
{Crawford}, C.~S., {Allen}, S.~W., {Ebeling}, H., {Edge}, A.~C., \& {Fabian}, A.~C. 1999, \mnras, 306, 857

\bibitem[{{de Plaa} {et~al.}(2017){de Plaa}, {Kaastra}, {Werner}, {Pinto}, {Kosec}, {Zhang}, {Mernier}, {Lovisari}, {Akamatsu}, {Schellenberger}, {Hofmann}, {Reiprich}, {Finoguenov}, {Ahoranta}, {Sanders}, {Fabian}, {Pols}, {Simionescu}, {Vink}, \& {B{\"o}hringer}}]{2017DePlaa}
{de Plaa}, J., {Kaastra}, J.~S., {Werner}, N., {et~al.} 2017, \aap, 607, A98

\bibitem[{{Donahue} \& {Voit}(2004)}]{2004Donahue}
{Donahue}, M. \& {Voit}, G.~M. 2004, in Clusters of Galaxies: Probes of Cosmological Structure and Galaxy Evolution, ed. J.~S. {Mulchaey}, A.~{Dressler}, \& A.~{Oemler}, 143

\bibitem[{{Ebeling} {et~al.}(1998){Ebeling}, {Edge}, {Bohringer}, {Allen}, {Crawford}, {Fabian}, {Voges}, \& {Huchra}}]{1998Ebeling}
{Ebeling}, H., {Edge}, A.~C., {Bohringer}, H., {et~al.} 1998, \mnras, 301, 881

\bibitem[{{Egami} {et~al.}(2006){Egami}, {Misselt}, {Rieke}, {Wise}, {Neugebauer}, {Kneib}, {Le Floc'h}, {Smith}, {Blaylock}, {Dole}, {Frayer}, {Huang}, {Krause}, {Papovich}, {P{\'e}rez-Gonz{\'a}lez}, \& {Rigby}}]{2006Egami}
{Egami}, E., {Misselt}, K.~A., {Rieke}, G.~H., {et~al.} 2006, \apj, 647, 922

\bibitem[{Ettori {et~al.}(2002)Ettori, Fabian, Allen, \& Johnstone}]{2002Ettori}
Ettori, S., Fabian, A.~C., Allen, S.~W., \& Johnstone, R.~M. 2002, Monthly Notices of the Royal Astronomical Society, 331, 635

\bibitem[{{Fabian}(1994)}]{1994Fabian}
{Fabian}, A.~C. 1994, \araa, 32, 277

\bibitem[{{Fabian}(2012)}]{2012Fabian}
{Fabian}, A.~C. 2012, \araa, 50, 455

\bibitem[{{Fabian} {et~al.}(2022){Fabian}, {Ferland}, {Sanders}, {McNamara}, {Pinto}, \& {Walker}}]{2022Fabian}
{Fabian}, A.~C., {Ferland}, G.~J., {Sanders}, J.~S., {et~al.} 2022, \mnras, 515, 3336

\bibitem[{{Fabian} \& {Nulsen}(1977)}]{1977Fabian}
{Fabian}, A.~C. \& {Nulsen}, P.~E.~J. 1977, \mnras, 180, 479

\bibitem[{{Fabian} {et~al.}(2011){Fabian}, {Sanders}, {Allen}, {Canning}, {Churazov}, {Crawford}, {Forman}, {Gabany}, {Hlavacek-Larrondo}, {Johnstone}, {Russell}, {Reynolds}, {Salom{\'e}}, {Taylor}, \& {Young}}]{2011Fabian}
{Fabian}, A.~C., {Sanders}, J.~S., {Allen}, S.~W., {et~al.} 2011, \mnras, 418, 2154

\bibitem[{{Fabian} {et~al.}(2003){Fabian}, {Sanders}, {Allen}, {Crawford}, {Iwasawa}, {Johnstone}, {Schmidt}, \& {Taylor}}]{2003Fabian}
{Fabian}, A.~C., {Sanders}, J.~S., {Allen}, S.~W., {et~al.} 2003, \mnras, 344, L43

\bibitem[{{Fabian} {et~al.}(2006){Fabian}, {Sanders}, {Taylor}, {Allen}, {Crawford}, {Johnstone}, \& {Iwasawa}}]{2006Fabian}
{Fabian}, A.~C., {Sanders}, J.~S., {Taylor}, G.~B., {et~al.} 2006, \mnras, 366, 417

\bibitem[{{Frank} {et~al.}(2013){Frank}, {Peterson}, {Andersson}, {Fabian}, \& {Sanders}}]{2013Frank}
{Frank}, K.~A., {Peterson}, J.~R., {Andersson}, K., {Fabian}, A.~C., \& {Sanders}, J.~S. 2013, \apj, 764, 46

\bibitem[{{Gaspari}(2015)}]{2015Gaspari}
{Gaspari}, M. 2015, \mnras, 451, L60

\bibitem[{{Gaspari} {et~al.}(2019){Gaspari}, {Eckert}, {Ettori}, {Tozzi}, {Bassini}, {Rasia}, {Brighenti}, {Sun}, {Borgani}, {Johnson}, {Tremblay}, {Stone}, {Temi}, {Yang}, {Tombesi}, \& {Cappi}}]{2019Gaspari}
{Gaspari}, M., {Eckert}, D., {Ettori}, S., {et~al.} 2019, \apj, 884, 169

\bibitem[{{Gaspari} {et~al.}(2018){Gaspari}, {McDonald}, {Hamer}, {Brighenti}, {Temi}, {Gendron-Marsolais}, {Hlavacek-Larrondo}, {Edge}, {Werner}, {Tozzi}, {Sun}, {Stone}, {Tremblay}, {Hogan}, {Eckert}, {Ettori}, {Yu}, {Biffi}, \& {Planelles}}]{2018Gaspari}
{Gaspari}, M., {McDonald}, M., {Hamer}, S.~L., {et~al.} 2018, \apj, 854, 167

\bibitem[{{Gaspari} {et~al.}(2012){Gaspari}, {Ruszkowski}, \& {Sharma}}]{2012Gaspari}
{Gaspari}, M., {Ruszkowski}, M., \& {Sharma}, P. 2012, \apj, 746, 94

\bibitem[{{Gaspari} {et~al.}(2020){Gaspari}, {Tombesi}, \& {Cappi}}]{2020Gaspari}
{Gaspari}, M., {Tombesi}, F., \& {Cappi}, M. 2020, Nature Astronomy, 4, 10

\bibitem[{{Gastaldello} {et~al.}(2021){Gastaldello}, {Simionescu}, {Mernier}, {Biffi}, {Gaspari}, {Sato}, \& {Matsushita}}]{2021Gastaldello}
{Gastaldello}, F., {Simionescu}, A., {Mernier}, F., {et~al.} 2021, Universe, 7, 208

\bibitem[{Giacintucci {et~al.}(2017)Giacintucci, Markevitch, Cassano, Venturi, Clarke, \& Brunetti}]{2017Giacintucci}
Giacintucci, S., Markevitch, M., Cassano, R., {et~al.} 2017, The Astrophysical Journal, 841, 71

\bibitem[{{Giacintucci} {et~al.}(2019){Giacintucci}, {Markevitch}, {Cassano}, {Venturi}, {Clarke}, {Kale}, \& {Cuciti}}]{2019Giacintucci}
{Giacintucci}, S., {Markevitch}, M., {Cassano}, R., {et~al.} 2019, \apj, 880, 70

\bibitem[{{Gilmour} {et~al.}(2009){Gilmour}, {Best}, \& {Almaini}}]{2009Gilmour}
{Gilmour}, R., {Best}, P., \& {Almaini}, O. 2009, \mnras, 392, 1509

\bibitem[{{Guainazzi} \& {Tashiro}(2019)}]{2019GuainazziTashiro}
{Guainazzi}, M. \& {Tashiro}, M. 2019, Proceedings of the IAU Symposium No. IAUS342

\bibitem[{{HI4PI Collaboration} {et~al.}(2016){HI4PI Collaboration}, {Ben Bekhti}, Fl{\"{o}}er, Keller, Kerp, Lenz, Winkel, Bailin, Calabretta, Dedes, Ford, Gibson, Haud, Janowiecki, Kalberla, Lockman, McClure-Griffiths, Murphy, Nakanishi, Pisano, \& Staveley-Smith}]{2016HI4PI}
{HI4PI Collaboration}, {Ben Bekhti}, N., Fl{\"{o}}er, L., {et~al.} 2016, A$\&$A, 594

\bibitem[{{Hitomi Collaboration} {et~al.}(2016){Hitomi Collaboration}, {Aharonian}, {Akamatsu}, {Akimoto}, {Allen}, {Anabuki}, {Angelini}, {Arnaud}, {Audard}, {Awaki}, {Axelsson}, {Bamba}, {Bautz}, {Blandford}, {Brenneman}, {Brown}, {Bulbul}, {Cackett}, {Chernyakova}, {Chiao}, {Coppi}, {Costantini}, {de Plaa}, {den Herder}, {Done}, {Dotani}, {Ebisawa}, {Eckart}, {Enoto}, {Ezoe}, {Fabian}, {Ferrigno}, {Foster}, {Fujimoto}, {Fukazawa}, {Furuzawa}, {Galeazzi}, {Gallo}, {Gandhi}, {Giustini}, {Goldwurm}, {Gu}, {Guainazzi}, {Haba}, {Hagino}, {Hamaguchi}, {Harrus}, {Hatsukade}, {Hayashi}, {Hayashi}, {Hayashida}, {Hiraga}, {Hornschemeier}, {Hoshino}, {Hughes}, {Iizuka}, {Inoue}, {Inoue}, {Ishibashi}, {Ishida}, {Ishikawa}, {Ishisaki}, {Itoh}, {Iyomoto}, {Kaastra}, {Kallman}, {Kamae}, {Kara}, {Kataoka}, {Katsuda}, {Katsuta}, {Kawaharada}, {Kawai}, {Kelley}, {Khangulyan}, {Kilbourne}, {King}, {Kitaguchi}, {Kitamoto}, {Kitayama}, {Kohmura}, {Kokubun}, {Koyama}, {Koyama}, {Kretschmar}, {Krimm}, {Kubota}, {Kunieda},
  {Laurent}, {Lebrun}, {Lee}, {Leutenegger}, {Limousin}, {Loewenstein}, {Long}, {Lumb}, {Madejski}, {Maeda}, {Maier}, {Makishima}, {Markevitch}, {Matsumoto}, {Matsushita}, {McCammon}, {McNamara}, {Mehdipour}, {Miller}, {Miller}, {Mineshige}, {Mitsuda}, {Mitsuishi}, {Miyazawa}, {Mizuno}, {Mori}, {Mori}, {Moseley}, {Mukai}, {Murakami}, {Murakami}, {Mushotzky}, {Nagino}, {Nakagawa}, {Nakajima}, {Nakamori}, {Nakano}, {Nakashima}, {Nakazawa}, {Nobukawa}, {Noda}, {Nomachi}, {O'Dell}, {Odaka}, {Ohashi}, {Ohno}, {Okajima}, {Ota}, {Ozaki}, {Paerels}, {Paltani}, {Parmar}, {Petre}, {Pinto}, {Pohl}, {Porter}, {Pottschmidt}, {Ramsey}, {Reynolds}, {Russell}, {Safi-Harb}, {Saito}, {Sakai}, {Sameshima}, {Sato}, {Sato}, {Sato}, {Sawada}, {Schartel}, {Serlemitsos}, {Seta}, {Shidatsu}, {Simionescu}, {Smith}, {Soong}, {Stawarz}, {Sugawara}, {Sugita}, {Szymkowiak}, {Tajima}, {Takahashi}, {Takahashi}, {Takeda}, {Takei}, {Tamagawa}, {Tamura}, {Tamura}, {Tanaka}, {Tanaka}, {Tanaka}, {Tashiro}, {Tawara}, {Terada}, {Terashima},
  {Tombesi}, {Tomida}, {Tsuboi}, {Tsujimoto}, {Tsunemi}, {Tsuru}, {Uchida}, {Uchiyama}, {Uchiyama}, {Ueda}, {Ueda}, {Ueno}, {Uno}, {Urry}, {Ursino}, {de Vries}, {Watanabe}, {Werner}, {Wik}, {Wilkins}, {Williams}, {Yamada}, {Yamaguchi}, {Yamaoka}, {Yamasaki}, {Yamauchi}, {Yamauchi}, {Yaqoob}, {Yatsu}, {Yonetoku}, {Yoshida}, {Yuasa}, {Zhuravleva}, \& {Zoghbi}}]{2016Hitomi}
{Hitomi Collaboration}, {Aharonian}, F., {Akamatsu}, H., {et~al.} 2016, \nat, 535, 117

\bibitem[{{Hudson} {et~al.}(2010){Hudson}, {Mittal}, {Reiprich}, {Nulsen}, {Andernach}, \& {Sarazin}}]{2010Hudson}
{Hudson}, D.~S., {Mittal}, R., {Reiprich}, T.~H., {et~al.} 2010, \aap, 513, A37

\bibitem[{{Iani} {et~al.}(2019){Iani}, {Rodighiero}, {Fritz}, {Cresci}, {Mancini}, {Tozzi}, {Rodr{\'\i}guez-Mu{\~n}oz}, {Rosati}, {Caminha}, {Zanella}, {Berta}, {Cassata}, {Concas}, {Enia}, {Fadda}, {Franceschini}, {Liu}, {Mercurio}, {Morselli}, {P{\'e}rez-Gonz{\'a}lez}, {Popesso}, {Sabatini}, {Vernet}, \& {van Weeren}}]{2019Iani}
{Iani}, E., {Rodighiero}, G., {Fritz}, J., {et~al.} 2019, \mnras, 487, 5593

\bibitem[{{Kaastra} {et~al.}(2001){Kaastra}, {Ferrigno}, {Tamura}, {Paerels}, {Peterson}, \& {Mittaz}}]{2001Kaastra}
{Kaastra}, J.~S., {Ferrigno}, C., {Tamura}, T., {et~al.} 2001, \aap, 365, L99

\bibitem[{{Kaastra} {et~al.}(1996){Kaastra}, {Mewe}, \& {Nieuwenhuijzen}}]{1996Kaastra}
{Kaastra}, J.~S., {Mewe}, R., \& {Nieuwenhuijzen}, H. 1996, in UV and X-ray Spectroscopy of Astrophysical and Laboratory Plasmas, 411--414

\bibitem[{{Kaastra} {et~al.}(2004){Kaastra}, {Tamura}, {Peterson}, {Bleeker}, {Ferrigno}, {Kahn}, {Paerels}, {Piffaretti}, {Branduardi-Raymont}, \& {B{\"o}hringer}}]{2004Kaastra}
{Kaastra}, J.~S., {Tamura}, T., {Peterson}, J.~R., {et~al.} 2004, \aap, 413, 415

\bibitem[{{Kadam} {et~al.}(2019){Kadam}, {Sonkamble}, {Pawar}, \& {Patil}}]{2019Kadam}
{Kadam}, S.~K., {Sonkamble}, S.~S., {Pawar}, P.~K., \& {Patil}, M.~K. 2019, \mnras, 484, 4113

\bibitem[{{Kale} {et~al.}(2015){Kale}, {Venturi}, {Cassano}, {Giacintucci}, {Bardelli}, {Dallacasa}, \& {Zucca}}]{2015Kale}
{Kale}, R., {Venturi}, T., {Cassano}, R., {et~al.} 2015, \aap, 581, A23

\bibitem[{{Knowles} {et~al.}(2022){Knowles}, {Cotton}, {Rudnick}, {Camilo}, {Goedhart}, {Deane}, {Ramatsoku}, {Bietenholz}, {Br{\"u}ggen}, {Button}, {Chen}, {Chibueze}, {Clarke}, {de Gasperin}, {Ianjamasimanana}, {J{\'o}zsa}, {Hilton}, {Kesebonye}, {Kolokythas}, {Kraan-Korteweg}, {Lawrie}, {Lochner}, {Loubser}, {Marchegiani}, {Mhlahlo}, {Moodley}, {Murphy}, {Namumba}, {Oozeer}, {Parekh}, {Pillay}, {Passmoor}, {Ramaila}, {Ranchod}, {Retana-Montenegro}, {Sebokolodi}, {Sikhosana}, {Smirnov}, {Thorat}, {Venturi}, {Abbott}, {Adam}, {Adams}, {Aldera}, {Bauermeister}, {Bennett}, {Bode}, {Botha}, {Botha}, {Brederode}, {Buchner}, {Burger}, {Cheetham}, {de Villiers}, {Dikgale-Mahlakoana}, {du Toit}, {Esterhuyse}, {Fadana}, {Fanaroff}, {Fataar}, {Foley}, {Fourie}, {Frank}, {Gamatham}, {Gatsi}, {Geyer}, {Gouws}, {Gumede}, {Heywood}, {Hlakola}, {Hokwana}, {Hoosen}, {Horn}, {Horrell}, {Hugo}, {Isaacson}, {Jonas}, {Jordaan}, {Joubert}, {Julie}, {Kapp}, {Kasper}, {Kenyon}, {Kotz{\'e}}, {Kotze}, {Kriek}, {Kriel}, {Krishnan},
  {Kusel}, {Legodi}, {Lehmensiek}, {Liebenberg}, {Lord}, {Lunsky}, {Madisa}, {Magnus}, {Main}, {Makhaba}, {Makhathini}, {Malan}, {Manley}, {Marais}, {Maree}, {Martens}, {Mauch}, {McAlpine}, {Merry}, {Millenaar}, {Mokone}, {Monama}, {Mphego}, {New}, {Ngcebetsha}, {Ngoasheng}, {Ockards}, {Otto}, {Patel}, {Peens-Hough}, {Perkins}, {Ramanujam}, {Ramudzuli}, {Ratcliffe}, {Renil}, {Robyntjies}, {Rust}, {Salie}, {Sambu}, {Schollar}, {Schwardt}, {Schwartz}, {Serylak}, {Siebrits}, {Sirothia}, {Slabber}, {Sofeya}, {Taljaard}, {Tasse}, {Tiplady}, {Toruvanda}, {Twum}, {van Balla}, {van der Byl}, {van der Merwe}, {van Dyk}, {Van Tonder}, {Van Wyk}, {Venter}, {Venter}, {Welz}, {Williams}, \& {Xaia}}]{2022Knowles}
{Knowles}, K., {Cotton}, W.~D., {Rudnick}, L., {et~al.} 2022, \aap, 657, A56

\bibitem[{{Liu} {et~al.}(2020){Liu}, {Tozzi}, {Ettori}, {De Grandi}, {Gastaldello}, {Rosati}, \& {Norman}}]{2020LiuA}
{Liu}, A., {Tozzi}, P., {Ettori}, S., {et~al.} 2020, \aap, 637, A58

\bibitem[{{Liu} {et~al.}(2019{\natexlab{a}}){Liu}, {Zhai}, \& {Tozzi}}]{2019LiuA}
{Liu}, A., {Zhai}, M., \& {Tozzi}, P. 2019{\natexlab{a}}, \mnras, 485, 1651

\bibitem[{{Liu} {et~al.}(2019{\natexlab{b}}){Liu}, {Pinto}, {Fabian}, {Russell}, \& {Sanders}}]{2019Liu}
{Liu}, H., {Pinto}, C., {Fabian}, A.~C., {Russell}, H.~R., \& {Sanders}, J.~S. 2019{\natexlab{b}}, \mnras, 485, 1757

\bibitem[{{Liu} {et~al.}(2024){Liu}, {Sun}, {Voit}, {Lal}, {Nulsen}, {Gaspari}, {Sarazin}, {Ehlert}, \& {Zheng}}]{2024Liu}
{Liu}, W., {Sun}, M., {Voit}, G.~M., {et~al.} 2024, \mnras, 531, 2063

\bibitem[{{Lodders} \& {Palme}(2009)}]{2009LoddersPalme}
{Lodders}, K. \& {Palme}, H. 2009, Meteoritics and Planetary Science Supplement, 72, 5154

\bibitem[{{Maccagni} {et~al.}(2021){Maccagni}, {Serra}, {Murgia}, {Govoni}, {Morokuma-Matsui}, \& {Kleiner}}]{2021Maccagni}
{Maccagni}, F.~M., {Serra}, P., {Murgia}, M., {et~al.} 2021, in Galaxy Evolution and Feedback across Different Environments, ed. T.~{Storchi Bergmann}, W.~{Forman}, R.~{Overzier}, \& R.~{Riffel}, Vol. 359, 141--146

\bibitem[{{Mann} \& {Ebeling}(2012)}]{2012Mann}
{Mann}, A.~W. \& {Ebeling}, H. 2012, \mnras, 420, 2120

\bibitem[{{Mantz} {et~al.}(2016){Mantz}, {Allen}, {Morris}, \& {Schmidt}}]{2016Mantz}
{Mantz}, A.~B., {Allen}, S.~W., {Morris}, R.~G., \& {Schmidt}, R.~W. 2016, \mnras, 456, 4020

\bibitem[{{McDonald} {et~al.}(2012){McDonald}, {Bayliss}, {Benson}, {Foley}, {Ruel}, {Sullivan}, {Veilleux}, {Aird}, {Ashby}, {Bautz}, {Bazin}, {Bleem}, {Brodwin}, {Carlstrom}, {Chang}, {Cho}, {Clocchiatti}, {Crawford}, {Crites}, {de Haan}, {Desai}, {Dobbs}, {Dudley}, {Egami}, {Forman}, {Garmire}, {George}, {Gladders}, {Gonzalez}, {Halverson}, {Harrington}, {High}, {Holder}, {Holzapfel}, {Hoover}, {Hrubes}, {Jones}, {Joy}, {Keisler}, {Knox}, {Lee}, {Leitch}, {Liu}, {Lueker}, {Luong-van}, {Mantz}, {Marrone}, {McMahon}, {Mehl}, {Meyer}, {Miller}, {Mocanu}, {Mohr}, {Montroy}, {Murray}, {Natoli}, {Padin}, {Plagge}, {Pryke}, {Rawle}, {Reichardt}, {Rest}, {Rex}, {Ruhl}, {Saliwanchik}, {Saro}, {Sayre}, {Schaffer}, {Shaw}, {Shirokoff}, {Simcoe}, {Song}, {Spieler}, {Stalder}, {Staniszewski}, {Stark}, {Story}, {Stubbs}, {{\v{S}}uhada}, {van Engelen}, {Vanderlinde}, {Vieira}, {Vikhlinin}, {Williamson}, {Zahn}, \& {Zenteno}}]{2012McDonald}
{McDonald}, M., {Bayliss}, M., {Benson}, B.~A., {et~al.} 2012, \nat, 488, 349

\bibitem[{{McDonald} {et~al.}(2018){McDonald}, {Gaspari}, {McNamara}, \& {Tremblay}}]{2018McDonald}
{McDonald}, M., {Gaspari}, M., {McNamara}, B.~R., \& {Tremblay}, G.~R. 2018, \apj, 858, 45

\bibitem[{{McDonald} {et~al.}(2015){McDonald}, {McNamara}, {van Weeren}, {Applegate}, {Bayliss}, {Bautz}, {Benson}, {Carlstrom}, {Bleem}, {Chatzikos}, {Edge}, {Fabian}, {Garmire}, {Hlavacek-Larrondo}, {Jones-Forman}, {Mantz}, {Miller}, {Stalder}, {Veilleux}, \& {ZuHone}}]{2015McDonald}
{McDonald}, M., {McNamara}, B.~R., {van Weeren}, R.~J., {et~al.} 2015, \apj, 811, 111

\bibitem[{{McDonald} {et~al.}(2019){McDonald}, {McNamara}, {Voit}, {Bayliss}, {Benson}, {Brodwin}, {Canning}, {Florian}, {Garmire}, {Gaspari}, {Gladders}, {Hlavacek-Larrondo}, {Kara}, {Reichardt}, {Russell}, {Saro}, {Sharon}, {Somboonpanyakul}, {Tremblay}, \& {van Weeren}}]{2019McDonald}
{McDonald}, M., {McNamara}, B.~R., {Voit}, G.~M., {et~al.} 2019, \apj, 885, 63

\bibitem[{{McKinley} {et~al.}(2022){McKinley}, {Tingay}, {Gaspari}, {Kraft}, {Matherne}, {Offringa}, {McDonald}, {Calzadilla}, {Veilleux}, {Shabala}, {Gwyn}, {Bland-Hawthorn}, {Crnojevi{\'c}}, {Gaensler}, \& {Johnston-Hollitt}}]{2022McKinley}
{McKinley}, B., {Tingay}, S.~J., {Gaspari}, M., {et~al.} 2022, Nature Astronomy, 6, 109

\bibitem[{{McNamara} \& {Nulsen}(2007)}]{2007McNamara}
{McNamara}, B.~R. \& {Nulsen}, P.~E.~J. 2007, \araa, 45, 117

\bibitem[{{McNamara} {et~al.}(2005){McNamara}, {Nulsen}, {Wise}, {Rafferty}, {Carilli}, {Sarazin}, \& {Blanton}}]{2005McNamara}
{McNamara}, B.~R., {Nulsen}, P.~E.~J., {Wise}, M.~W., {et~al.} 2005, \nat, 433, 45

\bibitem[{{McNamara} {et~al.}(2006){McNamara}, {Rafferty}, {B{\^\i}rzan}, {Steiner}, {Wise}, {Nulsen}, {Carilli}, {Ryan}, \& {Sharma}}]{2006McNamara}
{McNamara}, B.~R., {Rafferty}, D.~A., {B{\^\i}rzan}, L., {et~al.} 2006, \apj, 648, 164

\bibitem[{{McNamara} {et~al.}(2014){McNamara}, {Russell}, {Nulsen}, {Edge}, {Murray}, {Main}, {Vantyghem}, {Combes}, {Fabian}, {Salome}, {Kirkpatrick}, {Baum}, {Bregman}, {Donahue}, {Egami}, {Hamer}, {O'Dea}, {Oonk}, {Tremblay}, \& {Voit}}]{2014McNamara}
{McNamara}, B.~R., {Russell}, H.~R., {Nulsen}, P.~E.~J., {et~al.} 2014, arXiv e-prints, arXiv:1403.4249

\bibitem[{{McNamara} {et~al.}(2000){McNamara}, {Wise}, {Nulsen}, {David}, {Sarazin}, {Bautz}, {Markevitch}, {Vikhlinin}, {Forman}, {Jones}, \& {Harris}}]{2000McNamara}
{McNamara}, B.~R., {Wise}, M., {Nulsen}, P.~E.~J., {et~al.} 2000, \apjl, 534, L135

\bibitem[{{Merloni} {et~al.}(2003){Merloni}, {Heinz}, \& {di Matteo}}]{2003Merloni}
{Merloni}, A., {Heinz}, S., \& {di Matteo}, T. 2003, \mnras, 345, 1057

\bibitem[{{Mernier} {et~al.}(2017){Mernier}, {de Plaa}, {Kaastra}, {Zhang}, {Akamatsu}, {Gu}, {Kosec}, {Mao}, {Pinto}, {Reiprich}, {Sanders}, {Simionescu}, \& {Werner}}]{2017Mernier}
{Mernier}, F., {de Plaa}, J., {Kaastra}, J.~S., {et~al.} 2017, \aap, 603, A80

\bibitem[{{Mittal} {et~al.}(2017){Mittal}, {McDonald}, {Whelan}, \& {Bruzual}}]{2017Mittal}
{Mittal}, R., {McDonald}, M., {Whelan}, J.~T., \& {Bruzual}, G. 2017, \mnras, 465, 3143

\bibitem[{{Molendi} \& {Pizzolato}(2001)}]{2001MolendiPizzolato}
{Molendi}, S. \& {Pizzolato}, F. 2001, \apj, 560, 194

\bibitem[{{Molendi} {et~al.}(2016){Molendi}, {Tozzi}, {Gaspari}, {De Grandi}, {Gastaldello}, {Ghizzardi}, \& {Rossetti}}]{2016Molendi}
{Molendi}, S., {Tozzi}, P., {Gaspari}, M., {et~al.} 2016, \aap, 595, A123

\bibitem[{{Nandra} {et~al.}(2013){Nandra}, {Barret}, {Barcons}, {Fabian}, {den Herder}, {Piro}, {Watson}, {Adami}, {Aird}, {Afonso}, {Alexander}, {Argiroffi}, {Amati}, {Arnaud}, {Atteia}, {Audard}, {Badenes}, {Ballet}, {Ballo}, {Bamba}, {Bhardwaj}, {Stefano Battistelli}, {Becker}, {De Becker}, {Behar}, {Bianchi}, {Biffi}, {B{\^\i}rzan}, {Bocchino}, {Bogdanov}, {Boirin}, {Boller}, {Borgani}, {Borm}, {Bouch{\'e}}, {Bourdin}, {Bower}, {Braito}, {Branchini}, {Branduardi-Raymont}, {Bregman}, {Brenneman}, {Brightman}, {Br{\"u}ggen}, {Buchner}, {Bulbul}, {Brusa}, {Bursa}, {Caccianiga}, {Cackett}, {Campana}, {Cappelluti}, {Cappi}, {Carrera}, {Ceballos}, {Christensen}, {Chu}, {Churazov}, {Clerc}, {Corbel}, {Corral}, {Comastri}, {Costantini}, {Croston}, {Dadina}, {D'Ai}, {Decourchelle}, {Della Ceca}, {Dennerl}, {Dolag}, {Done}, {Dovciak}, {Drake}, {Eckert}, {Edge}, {Ettori}, {Ezoe}, {Feigelson}, {Fender}, {Feruglio}, {Finoguenov}, {Fiore}, {Galeazzi}, {Gallagher}, {Gandhi}, {Gaspari}, {Gastaldello}, {Georgakakis},
  {Georgantopoulos}, {Gilfanov}, {Gitti}, {Gladstone}, {Goosmann}, {Gosset}, {Grosso}, {Guedel}, {Guerrero}, {Haberl}, {Hardcastle}, {Heinz}, {Alonso Herrero}, {Herv{\'e}}, {Holmstrom}, {Iwasawa}, {Jonker}, {Kaastra}, {Kara}, {Karas}, {Kastner}, {King}, {Kosenko}, {Koutroumpa}, {Kraft}, {Kreykenbohm}, {Lallement}, {Lanzuisi}, {Lee}, {Lemoine-Goumard}, {Lobban}, {Lodato}, {Lovisari}, {Lotti}, {McCharthy}, {McNamara}, {Maggio}, {Maiolino}, {De Marco}, {de Martino}, {Mateos}, {Matt}, {Maughan}, {Mazzotta}, {Mendez}, {Merloni}, {Micela}, {Miceli}, {Mignani}, {Miller}, {Miniutti}, {Molendi}, {Montez}, {Moretti}, {Motch}, {Naz{\'e}}, {Nevalainen}, {Nicastro}, {Nulsen}, {Ohashi}, {O'Brien}, {Osborne}, {Oskinova}, {Pacaud}, {Paerels}, {Page}, {Papadakis}, {Pareschi}, {Petre}, {Petrucci}, {Piconcelli}, {Pillitteri}, {Pinto}, {de Plaa}, {Pointecouteau}, {Ponman}, {Ponti}, {Porquet}, {Pounds}, {Pratt}, {Predehl}, {Proga}, {Psaltis}, {Rafferty}, {Ramos-Ceja}, {Ranalli}, {Rasia}, {Rau}, {Rauw}, {Rea}, {Read}, {Reeves},
  {Reiprich}, {Renaud}, {Reynolds}, {Risaliti}, {Rodriguez}, {Rodriguez Hidalgo}, {Roncarelli}, {Rosario}, {Rossetti}, {Rozanska}, {Rovilos}, {Salvaterra}, {Salvato}, {Di Salvo}, {Sanders}, {Sanz-Forcada}, {Schawinski}, {Schaye}, {Schwope}, {Sciortino}, {Severgnini}, {Shankar}, {Sijacki}, {Sim}, {Schmid}, {Smith}, {Steiner}, {Stelzer}, {Stewart}, {Strohmayer}, {Str{\"u}der}, {Sun}, {Takei}, {Tatischeff}, {Tiengo}, {Tombesi}, {Trinchieri}, {Tsuru}, {Ud-Doula}, {Ursino}, {Valencic}, {Vanzella}, {Vaughan}, {Vignali}, {Vink}, {Vito}, {Volonteri}, {Wang}, {Webb}, {Willingale}, {Wilms}, {Wise}, {Worrall}, {Young}, {Zampieri}, {In't Zand}, {Zane}, {Zezas}, {Zhang}, \& {Zhuravleva}}]{2013Nandra}
{Nandra}, K., {Barret}, D., {Barcons}, X., {et~al.} 2013, arXiv e-prints, arXiv:1306.2307

\bibitem[{{North} {et~al.}(2021){North}, {Davis}, {Bureau}, {Gaspari}, {Cappellari}, {Iguchi}, {Liu}, {Onishi}, {Sarzi}, {Smith}, \& {Williams}}]{2021North}
{North}, E.~V., {Davis}, T.~A., {Bureau}, M., {et~al.} 2021, \mnras, 503, 5179

\bibitem[{{Olivares} {et~al.}(2022){Olivares}, {Salom{\'e}}, {Hamer}, {Combes}, {Gaspari}, {Kolokythas}, {O'Sullivan}, {Beckmann}, {Babul}, {Polles}, {Lehnert}, {Loubser}, {Donahue}, {Gendron-Marsolais}, {Lagos}, {Pineau des Forets}, {Godard}, {Rose}, {Tremblay}, {Ferland}, \& {Guillard}}]{2022Olivares}
{Olivares}, V., {Salom{\'e}}, P., {Hamer}, S.~L., {et~al.} 2022, \aap, 666, A94

\bibitem[{{Pasini} {et~al.}(2021){Pasini}, {Gitti}, {Brighenti}, {O'Sullivan}, {Gastaldello}, {Temi}, \& {Hamer}}]{2021Pasini}
{Pasini}, T., {Gitti}, M., {Brighenti}, F., {et~al.} 2021, \apj, 911, 66

\bibitem[{{Peterson} \& {Fabian}(2006)}]{2006Peterson}
{Peterson}, J.~R. \& {Fabian}, A.~C. 2006, \physrep, 427, 1

\bibitem[{{Peterson} {et~al.}(2001){Peterson}, {Paerels}, {Kaastra}, {Arnaud}, {Reiprich}, {Fabian}, {Mushotzky}, {Jernigan}, \& {Sakelliou}}]{2001Peterson}
{Peterson}, J.~R., {Paerels}, F.~B.~S., {Kaastra}, J.~S., {et~al.} 2001, \aap, 365, L104

\bibitem[{{Phipps} {et~al.}(2019){Phipps}, {Bogd{\'a}n}, {Lovisari}, {Kov{\'a}cs}, {Volonteri}, \& {Dubois}}]{2019Phipps}
{Phipps}, F., {Bogd{\'a}n}, {\'A}., {Lovisari}, L., {et~al.} 2019, \apj, 875, 141

\bibitem[{{Pinto} {et~al.}(2018){Pinto}, {Bambic}, {Sanders}, {Fabian}, {McDonald}, {Russell}, {Liu}, \& {Reynolds}}]{2018Pinto}
{Pinto}, C., {Bambic}, C.~J., {Sanders}, J.~S., {et~al.} 2018, \mnras, 480, 4113

\bibitem[{{Pinto} {et~al.}(2015){Pinto}, {Sanders}, {Werner}, {de Plaa}, {Fabian}, {Zhang}, {Kaastra}, {Finoguenov}, \& {Ahoranta}}]{2015Pinto}
{Pinto}, C., {Sanders}, J.~S., {Werner}, N., {et~al.} 2015, \aap, 575, A38

\bibitem[{{Planck Collaboration} {et~al.}(2020){Planck Collaboration}, Aghanim, Akrami, Ashdown, Aumont, Baccigalupi, Ballardini, Banday, Barreiro, Bartolo, Basak, Battye, Benabed, Bernard, Bersanelli, Bielewicz, Bock, Bond, Borrill, Bouchet, Boulanger, Bucher, Burigana, Butler, Calabrese, Cardoso, Carron, Challinor, Chiang, Chluba, Colombo, Combet, Contreras, Crill, Cuttaia, de~Bernardis, de~Zotti, Delabrouille, Delouis, {Di Valentino}, Diego, Dor{\'{e}}, Douspis, Ducout, Dupac, Dusini, Efstathiou, Elsner, En{\ss}lin, Eriksen, Fantaye, Farhang, Fergusson, Fernandez-Cobos, Finelli, Forastieri, Frailis, Fraisse, Franceschi, Frolov, Galeotta, Galli, Ganga, G{\'{e}}nova-Santos, Gerbino, Ghosh, Gonz{\'{a}}lez-Nuevo, G{\'{o}}rski, Gratton, Gruppuso, Gudmundsson, Hamann, Handley, Hansen, Herranz, Hildebrandt, Hivon, Huang, Jaffe, Jones, Karakci, Keih{\"{a}}nen, Keskitalo, Kiiveri, Kim, Kisner, Knox, Krachmalnicoff, Kunz, Kurki-Suonio, Lagache, Lamarre, Lasenby, Lattanzi, Lawrence, {Le Jeune}, Lemos, Lesgourgues,
  Levrier, Lewis, Liguori, Lilje, Lilley, Lindholm, L{\'{o}}pez-Caniego, Lubin, Ma, Mac\'ias-P{\'{e}}rez, Maggio, Maino, Mandolesi, Mangilli, Marcos-Caballero, Maris, Martin, Martinelli, Mart\'inez-Gonz{\'{a}}lez, Matarrese, Mauri, McEwen, Meinhold, Melchiorri, Mennella, Migliaccio, Millea, Mitra, Miville-Desch{\^{e}}nes, Molinari, Montier, Morgante, Moss, Natoli, N{\o}rgaard-Nielsen, Pagano, Paoletti, Partridge, Patanchon, Peiris, Perrotta, Pettorino, Piacentini, Polastri, Polenta, Puget, Rachen, Reinecke, Remazeilles, Renzi, Rocha, Rosset, Roudier, Rubi{\~{n}}o-Mart\'in, Ruiz-Granados, Salvati, Sandri, Savelainen, Scott, Shellard, Sirignano, Sirri, Spencer, Sunyaev, Suur-Uski, Tauber, Tavagnacco, Tenti, Toffolatti, Tomasi, Trombetti, Valenziano, Valiviita, {Van Tent}, Vibert, Vielva, Villa, Vittorio, Wandelt, Wehus, White, White, Zacchei, \& Zonca}]{2020Planck}
{Planck Collaboration}, Aghanim, N., Akrami, Y., {et~al.} 2020, Astronomy and Astrophysics, 641, A6

\bibitem[{{Rawle} {et~al.}(2012){Rawle}, {Edge}, {Egami}, {Rex}, {Smith}, {Altieri}, {Fiedler}, {Haines}, {Pereira}, {P{\'e}rez-Gonz{\'a}lez}, {Portouw}, {Valtchanov}, {Walth}, {van der Werf}, \& {Zemcov}}]{2012Rawle}
{Rawle}, T.~D., {Edge}, A.~C., {Egami}, E., {et~al.} 2012, \apj, 747, 29

\bibitem[{{Rizza} {et~al.}(1998){Rizza}, {Burns}, {Ledlow}, {Owen}, {Voges}, \& {Bliton}}]{1998Rizza}
{Rizza}, E., {Burns}, J.~O., {Ledlow}, M.~J., {et~al.} 1998, \mnras, 301, 328

\bibitem[{Russell {et~al.}(2015)Russell, Fabian, McNamara, \& Broderick}]{2015Russell}
Russell, H.~R., Fabian, A.~C., McNamara, B.~R., \& Broderick, A.~E. 2015, Monthly Notices of the Royal Astronomical Society, 451, 588

\bibitem[{{Sanders} \& {Fabian}(2007)}]{2007Sanders}
{Sanders}, J.~S. \& {Fabian}, A.~C. 2007, \mnras, 381, 1381

\bibitem[{{Sanders} \& {Fabian}(2013)}]{2013Sanders}
{Sanders}, J.~S. \& {Fabian}, A.~C. 2013, \mnras, 429, 2727

\bibitem[{{Sanders} {et~al.}(2010){Sanders}, {Fabian}, {Smith}, \& {Peterson}}]{2010Sanders}
{Sanders}, J.~S., {Fabian}, A.~C., {Smith}, R.~K., \& {Peterson}, J.~R. 2010, \mnras, 402, L11

\bibitem[{Sanders {et~al.}(2014)Sanders, Fabian, Sun, Churazov, Simionescu, Walker, \& Werner}]{2014Sanders}
Sanders, J.~S., Fabian, A.~C., Sun, M., {et~al.} 2014, Monthly Notices of the Royal Astronomical Society, 439, 1182

\bibitem[{{Santos} {et~al.}(2010){Santos}, {Tozzi}, {Rosati}, \& {B{\"o}hringer}}]{2010Santos}
{Santos}, J.~S., {Tozzi}, P., {Rosati}, P., \& {B{\"o}hringer}, H. 2010, \aap, 521, A64

\bibitem[{{Sonkamble} {et~al.}(2024){Sonkamble}, {Kadam}, {Paul}, {Pandge}, {Pawar}, \& {Patil}}]{2024Sonkamble}
{Sonkamble}, S.~S., {Kadam}, S.~K., {Paul}, S., {et~al.} 2024, Journal of Astrophysics and Astronomy, 45, 23

\bibitem[{{Takahashi} {et~al.}(2010){Takahashi}, {Mitsuda}, {Kelley}, {Aharonian}, {Akimoto}, {Allen}, {Anabuki}, {Angelini}, {Arnaud}, {Awaki}, {Bamba}, {Bando}, {Bautz}, {Blandford}, {Boyce}, {Brown}, {Chernyakova}, {Coppi}, {Costantini}, {Cottam}, {Crow}, {de Plaa}, {de Vries}, {den Herder}, {Dipirro}, {Done}, {Dotani}, {Ebisawa}, {Enoto}, {Ezoe}, {Fabian}, {Fujimoto}, {Fukazawa}, {Funk}, {Furuzawa}, {Galeazzi}, {Gandhi}, {Gendreau}, {Gilmore}, {Haba}, {Hamaguchi}, {Hatsukade}, {Hayashida}, {Hiraga}, {Hirose}, {Hornschemeier}, {Hughes}, {Hwang}, {Iizuka}, {Ishibashi}, {Ishida}, {Ishimura}, {Ishisaki}, {Isobe}, {Ito}, {Iwata}, {Kaastra}, {Kallman}, {Kamae}, {Katagiri}, {Kataoka}, {Katsuda}, {Kawaharada}, {Kawai}, {Kawasaki}, {Khangaluyan}, {Kilbourne}, {Kinugasa}, {Kitamoto}, {Kitayama}, {Kohmura}, {Kokubun}, {Kosaka}, {Kotani}, {Koyama}, {Kubota}, {Kunieda}, {Laurent}, {Lebrun}, {Limousin}, {Loewenstein}, {Long}, {Madejski}, {Maeda}, {Makishima}, {Markevitch}, {Matsumoto}, {Matsushita}, {McCammon},
  {Miller}, {Mineshige}, {Minesugi}, {Miyazawa}, {Mizuno}, {Mori}, {Mori}, {Mukai}, {Murakami}, {Murakami}, {Mushotzky}, {Nakagawa}, {Nakagawa}, {Nakajima}, {Nakamori}, {Nakazawa}, {Namba}, {Nomachi}, {O'Dell}, {Ogawa}, {Ogawa}, {Ogi}, {Ohashi}, {Ohno}, {Ohta}, {Okajima}, {Ota}, {Ozaki}, {Paerels}, {Paltani}, {Parmar}, {Petre}, {Pohl}, {Porter}, {Ramsey}, {Reynolds}, {Sakai}, {Sambruna}, {Sato}, {Sato}, {Serlemitsos}, {Shida}, {Shimada}, {Shinozaki}, {Shirron}, {Smith}, {Sneiderman}, {Soong}, {Stawarz}, {Sugita}, {Szymkowiak}, {Tajima}, {Takahashi}, {Takei}, {Tamagawa}, {Tamura}, {Tamura}, {Tanaka}, {Tanaka}, {Tanaka}, {Tashiro}, {Tawara}, {Terada}, {Terashima}, {Tombesi}, {Tomida}, {Tozuka}, {Tsuboi}, {Tsujimoto}, {Tsunemi}, {Tsuru}, {Uchida}, {Uchiyama}, {Uchiyama}, {Ueda}, {Uno}, {Urry}, {Watanabe}, {White}, {Yamada}, {Yamaguchi}, {Yamaoka}, {Yamasaki}, {Yamauchi}, {Yamauchi}, {Yatsu}, {Yonetoku}, \& {Yoshida}}]{2010Takahashi}
{Takahashi}, T., {Mitsuda}, K., {Kelley}, R., {et~al.} 2010, in Society of Photo-Optical Instrumentation Engineers (SPIE) Conference Series, Vol. 7732, Space Telescopes and Instrumentation 2010: Ultraviolet to Gamma Ray, ed. M.~{Arnaud}, S.~S. {Murray}, \& T.~{Takahashi}, 77320Z

\bibitem[{{Tamura} {et~al.}(2001){Tamura}, {Kaastra}, {Peterson}, {Paerels}, {Mittaz}, {Trudolyubov}, {Stewart}, {Fabian}, {Mushotzky}, {Lumb}, \& {Ikebe}}]{2001Tamura}
{Tamura}, T., {Kaastra}, J.~S., {Peterson}, J.~R., {et~al.} 2001, \aap, 365, L87

\bibitem[{{Timmerman} {et~al.}(2022){Timmerman}, {van Weeren}, {Botteon}, {R{\"o}ttgering}, {McNamara}, {Sweijen}, {B{\^\i}rzan}, \& {Morabito}}]{2022Timmerman}
{Timmerman}, R., {van Weeren}, R.~J., {Botteon}, A., {et~al.} 2022, \aap, 668, A65

\bibitem[{{Tozzi} {et~al.}(2015){Tozzi}, {Gastaldello}, {Molendi}, {Ettori}, {Santos}, {De Grandi}, {Balestra}, {Rosati}, {Altieri}, {Cresci}, {Menanteau}, \& {Valtchanov}}]{2015Tozzi}
{Tozzi}, P., {Gastaldello}, F., {Molendi}, S., {et~al.} 2015, \aap, 580, A6

\bibitem[{{van Weeren} {et~al.}(2014){van Weeren}, {Intema}, {Lal}, {Andrade-Santos}, {Br{\"u}ggen}, {de Gasperin}, {Forman}, {Hoeft}, {Jones}, {Nuza}, {R{\"o}ttgering}, \& {Stroe}}]{2014vanWeeren}
{van Weeren}, R.~J., {Intema}, H.~T., {Lal}, D.~V., {et~al.} 2014, \apjl, 786, L17

\bibitem[{Vantyghem {et~al.}(2014)Vantyghem, McNamara, Russell, Main, Nulsen, Wise, Hoekstra, \& Gitti}]{2014Vantyghem}
Vantyghem, A.~N., McNamara, B.~R., Russell, H.~R., {et~al.} 2014, Monthly Notices of the Royal Astronomical Society, 442, 3192

\bibitem[{{Wang} {et~al.}(2023){Wang}, {Tozzi}, {Yu}, {Gaspari}, \& {Ettori}}]{2023Wang}
{Wang}, L., {Tozzi}, P., {Yu}, H., {Gaspari}, M., \& {Ettori}, S. 2023, \aap, 674, A102

\bibitem[{Werner {et~al.}(2012)Werner, Allen, \& Simionescu}]{2012Werner}
Werner, N., Allen, S.~W., \& Simionescu, A. 2012, Monthly Notices of the Royal Astronomical Society, 425, 2731

\bibitem[{{Wittor} \& {Gaspari}(2020)}]{2020Wittor}
{Wittor}, D. \& {Gaspari}, M. 2020, \mnras, 498, 4983

\bibitem[{{Wittor} \& {Gaspari}(2023)}]{2023Wittor}
{Wittor}, D. \& {Gaspari}, M. 2023, \mnras, 521, L79

\bibitem[{Xu {et~al.}(2002)Xu, Kahn, Peterson, Behar, Paerels, Mushotzky, Jernigan, Brinkman, \& Makishima}]{2002Xu}
Xu, H., Kahn, S.~M., Peterson, J.~R., {et~al.} 2002, The Astrophysical Journal, 579, 600

\bibitem[{{Yang} {et~al.}(2018){Yang}, {Tozzi}, {Yu}, {Lusso}, {Gaspari}, {Gilli}, {Nardini}, \& {Risaliti}}]{2018Yang}
{Yang}, L., {Tozzi}, P., {Yu}, H., {et~al.} 2018, \apj, 859, 65

\bibitem[{{Zhang} {et~al.}(2016){Zhang}, {Xu}, {Zhu}, {Li}, {Hu}, {Wang}, {Gu}, {Gu}, {Zhang}, {Liu}, {Zhu}, \& {Wu}}]{2016Zhang}
{Zhang}, C., {Xu}, H., {Zhu}, Z., {et~al.} 2016, \apj, 823, 116

\end{thebibliography}

\newpage

\begin{appendix}

\section{Impact of active galactic nuclei variability on the fitting procedure}
\label{app:AGN variability}

In this work, we modeled the AGN component in the fitting process using a power law with the intrinsic absorption, $\rm N_{H}$, photon index, $\Gamma$, and luminosity, $\rm L_{X}$, fixed to the best-fit values obtained in \cite{2018Yang}. In addition, we did not account for AGN variability. However, the values for $\rm N_{H}$ and luminosity have rather large statistical uncertainties: $\rm N_{H}(10^{22}cm^{-2})=13.7^{+7.6}_{-4.9}$ and $\rm log(L_{X})[erg/s]=43.4^{+0.2}_{-0.4}$. In this appendix, we have performed additional tests and seen if there are some variations in the parameters relevant to our science goals associated with uncertainties in the AGN modelization. We thus allowed the intrinsic absorption and luminosity to vary within their respective $1\sigma$ error bounds, keeping $\Gamma$ fixed to 1.8. 

The results of these tests show no substantial differences in the fit parameters, with variations limited to within 5\% in most of the best-fit values. We further extended the analysis by considering $2 \sigma$ and $3 \sigma$ deviations. Also, in this case, the variations are limited to 5\% at most. These findings suggest that the variability of the AGN parameters within the observed error bounds does not affect the conclusions of this paper.

\section{EPIC spectra}
\label{app:EPIC spectra}

\begin{figure*}
   \centering
   \resizebox{!}{0.49\textwidth}{\includegraphics{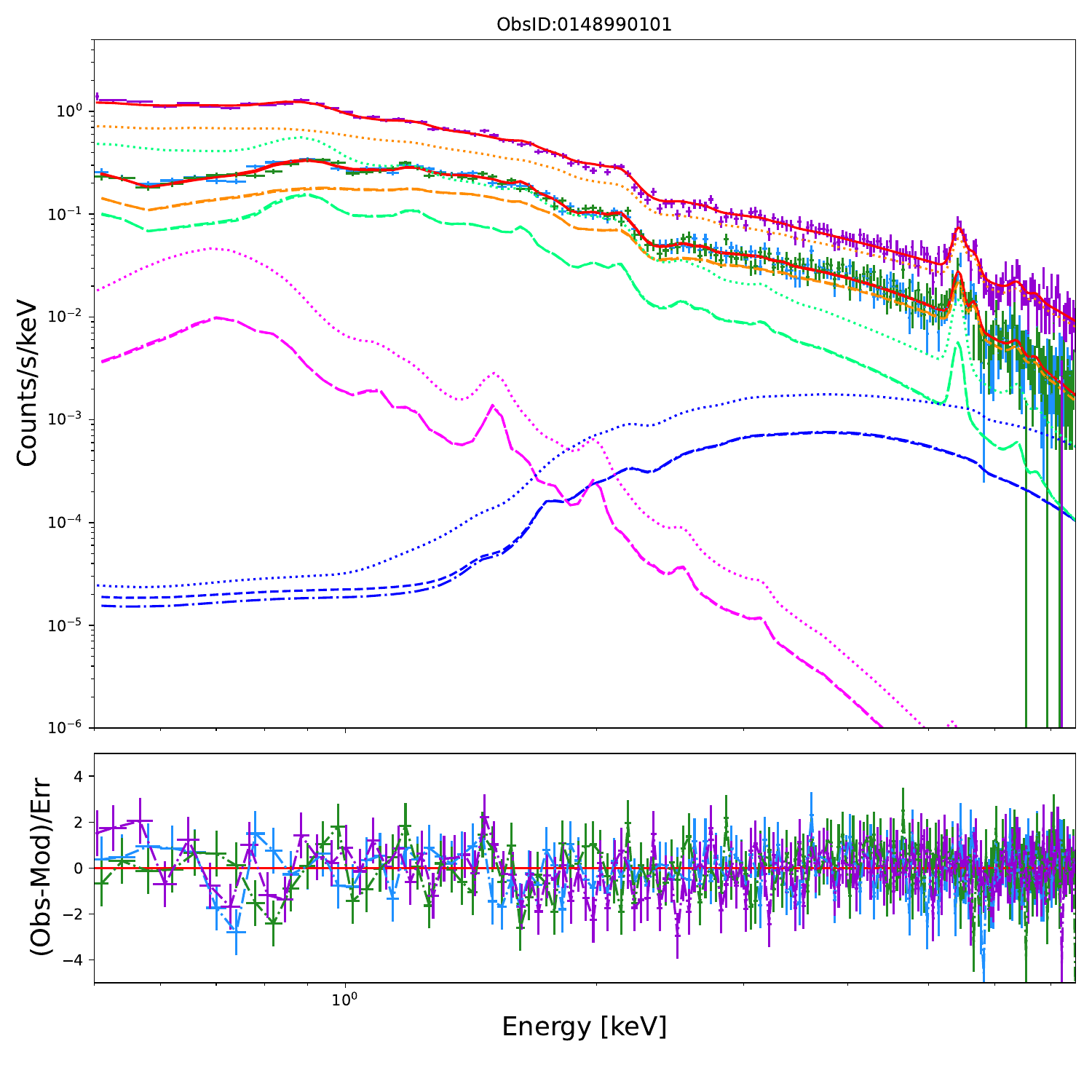}}
   \resizebox{!}{0.49\textwidth}{\includegraphics{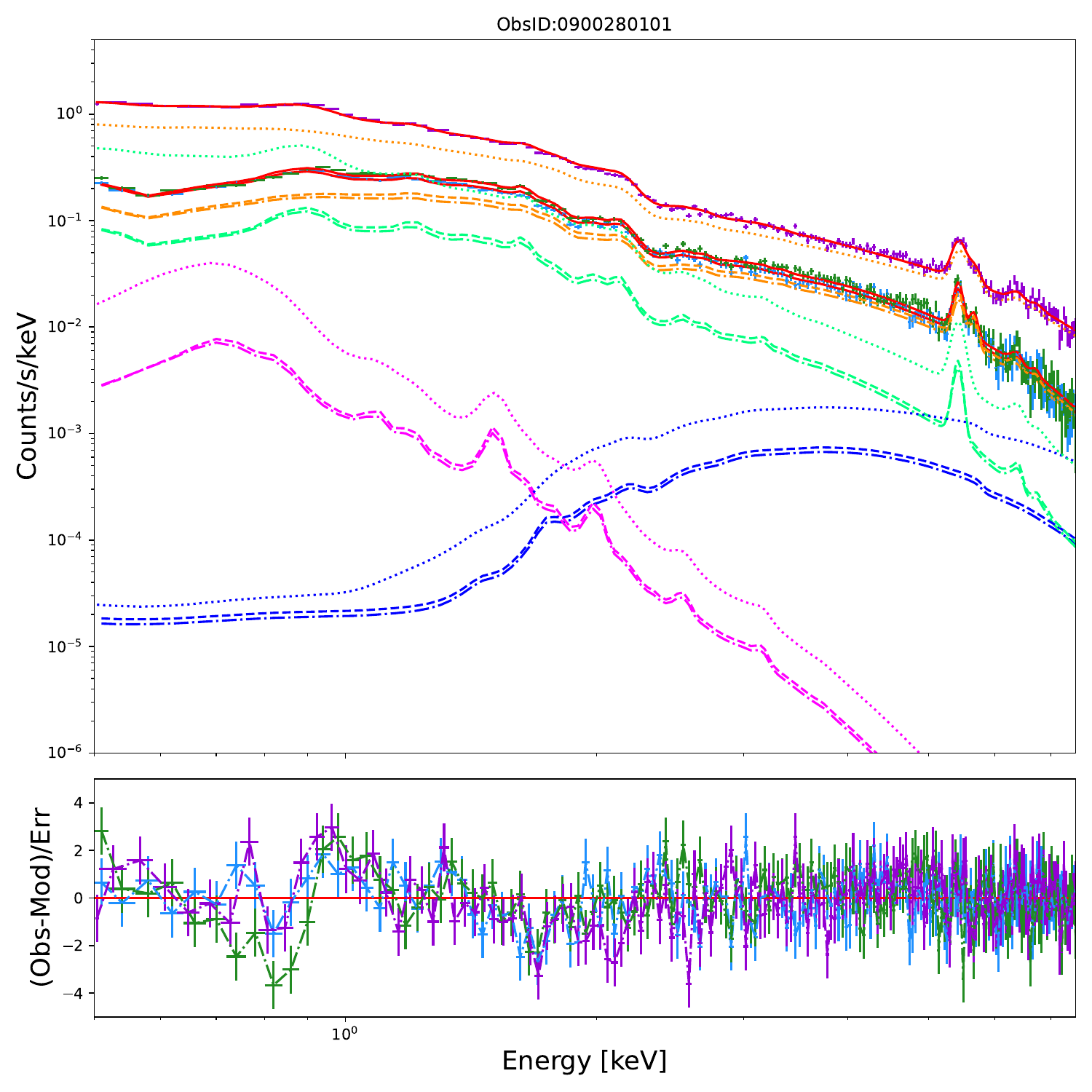}}
   \resizebox{!}{0.49\textwidth}{\includegraphics{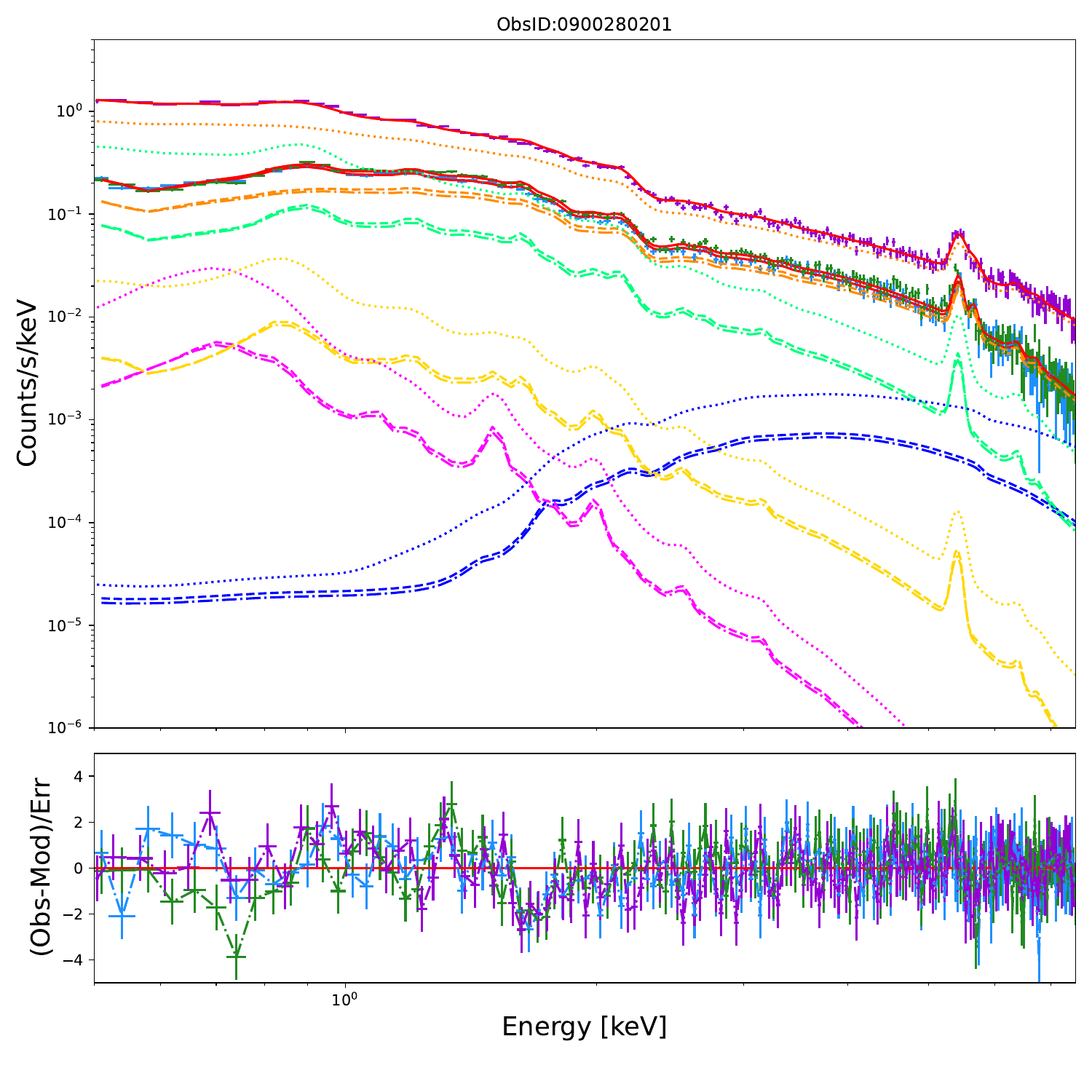}}
   \resizebox{!}{0.49\textwidth}{\includegraphics{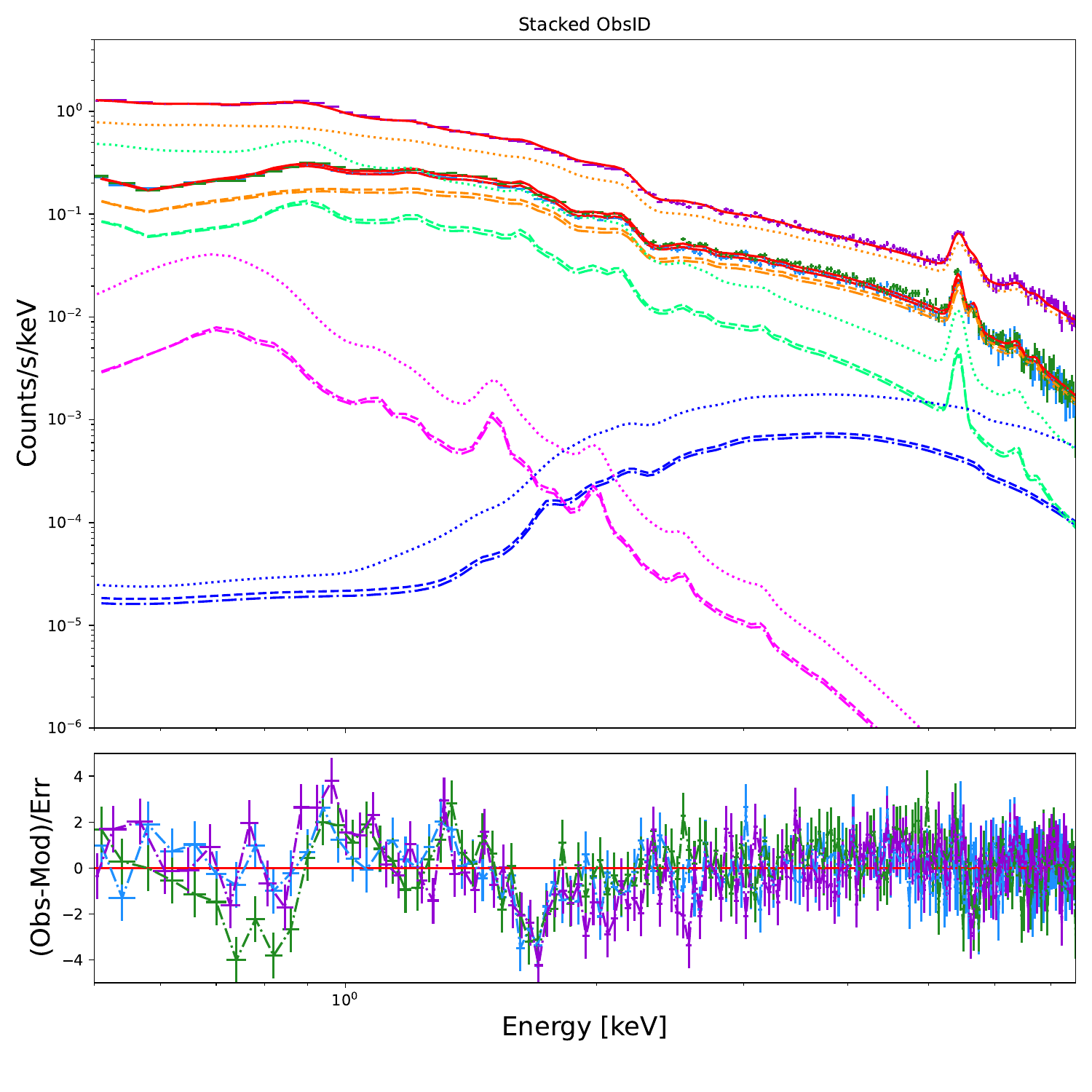}}
   \resizebox{!}{0.095\textwidth}{\includegraphics{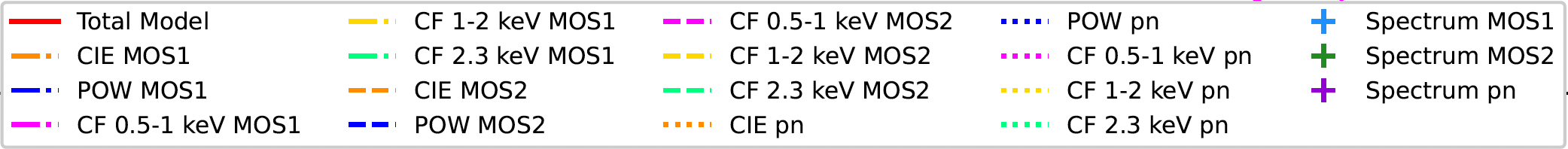}}
   \caption[A2667 Spectral fit of MOS1+MOS2 and pn data for ObsID 0148990101, 0900280101, 0900280201, and all stacked spectra]{Spectral fit of MOS1+MOS2 (light blue and green points) and pn (purple points) data for ObsID 0148990101 (upper left panel), 0900280101 (upper right panel), 0900280201 (lower left panel) and all stacked spectra (lower right panel). Residuals are shown in the bottom panel. 
   The model used for fitting is the combination of \emph{cie} and the isobaric CF split into three temperature intervals. The model components are represented with different colors and line styles in the plots. Orange lines represent the \emph{cie} component, and blue lines represent the power law component describing the ICM and AGN, respectively. Magenta, gold, and light green lines represent the three temperature bins (0.5-1, 1-2, and 2-3 keV) components.  
   }
    \label{fig:ObsID spectral fitting 1}
\end{figure*}

In this section we present the best-fit model plots of the stacked EPIC MOS1, MOS2, and pn spectra, along with the joint fit of all datasets. The spectra and corresponding best-fit models are shown in Fig. \ref{fig:ObsID spectral fitting 1}, where we also display each model component: the power-law and \emph{cie} components representing the ICM and AGN, as well as the three temperature bins of the CF component. The detailed fitting results are provided in Table \ref{tab:results EPIC fit}.

The plots reveal some residuals below 2 keV, particularly in the spectra from ObsID 0900280101 and 0900280201, which are the newly added observations. In contrast, these residuals are not evident in the spectrum from the older ObsID 0148990101. The discrepancies observed in the new data may be attributed to the higher signal-to-noise ratio of these new deeper observations compared to the older dataset. This increased sensitivity can amplify the impact of systematic uncertainties. Indeed, there are known calibration uncertainties of 5-10\%\footnote{https://xmmweb.esac.esa.int/docs/documents/CAL-TN-0018.pdf} in EPIC/MOS and EPIC/pn data, which may contribute to the observed residuals.
To address this issue, we also calculate the residuals in percentages below 2 keV. We found that they are consistent with 10\%.

It is also worth noting that the current model may not fully capture the features of the new observations, particularly the emission lines below 1 keV. This suggests that further refinement of the model, for example with additional local fitting procedures (as in \citealp{2017Mernier}) or additional components, might be necessary to better characterize these data.

\section{EPIC and RGS joint fit spectra}
\label{app:EPIC and RGS joint fit spectra}

\begin{figure*}
   \centering
   \resizebox{!}{0.55\textwidth}{\includegraphics[scale=0.3]{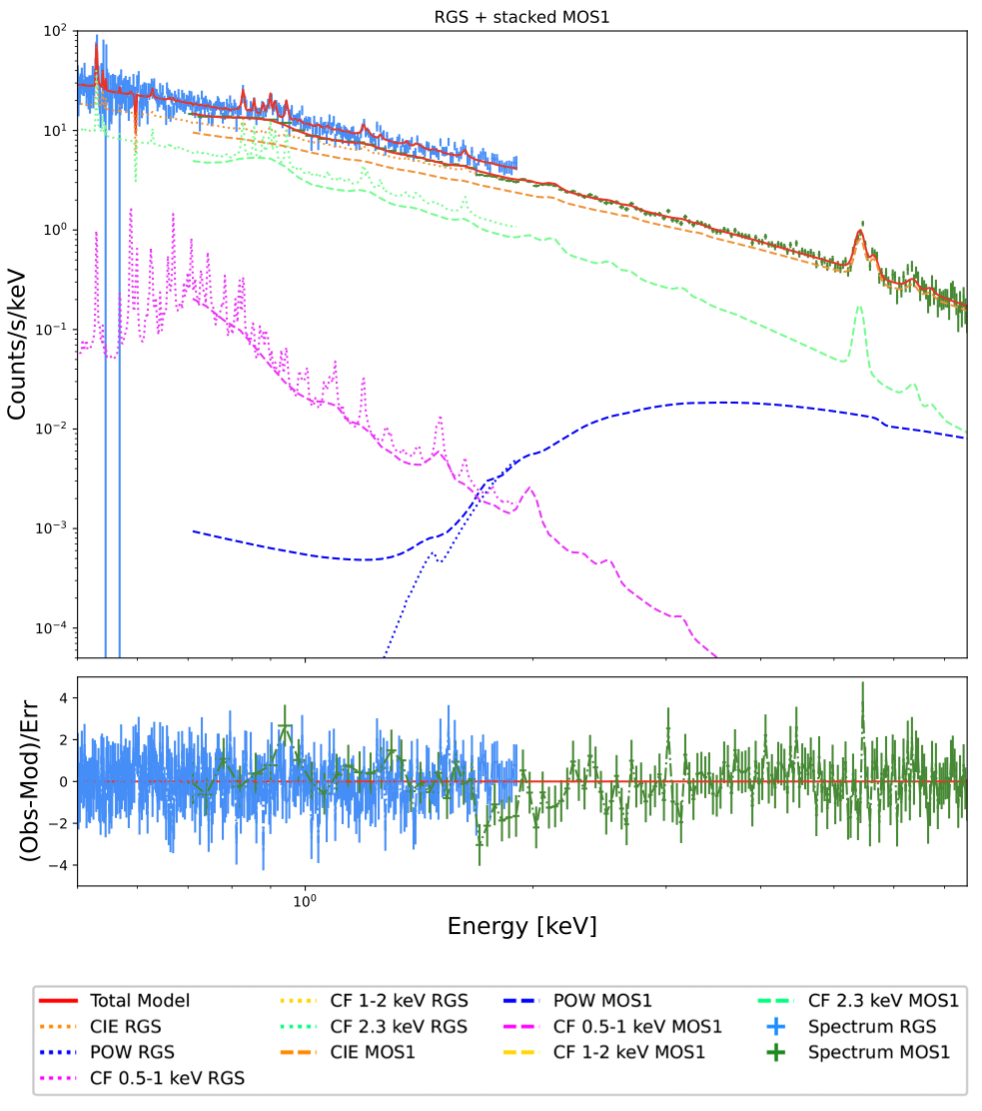}}
   \resizebox{!}{0.55\textwidth}{\includegraphics[scale=0.3]{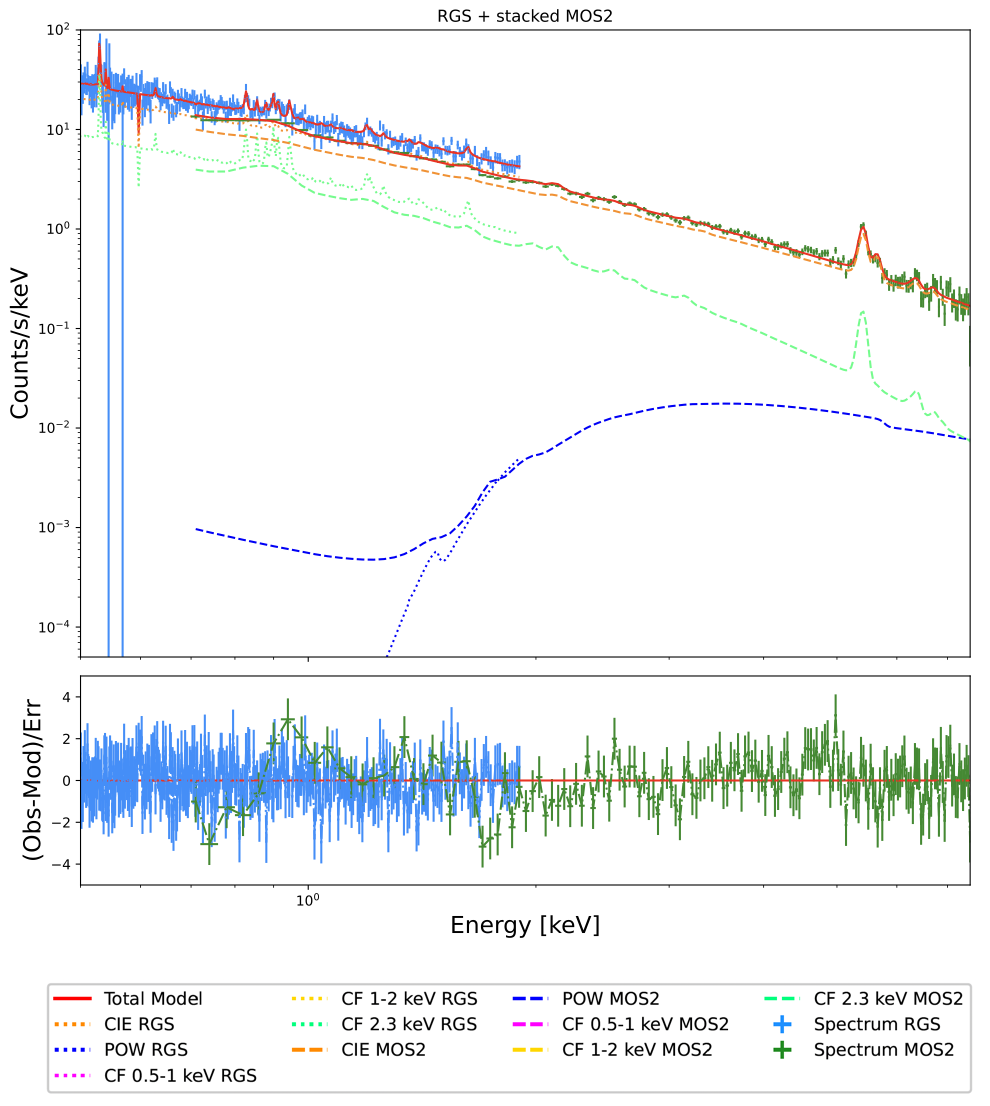}}
   \resizebox{!}{0.55\textwidth}{\includegraphics[scale=0.3]{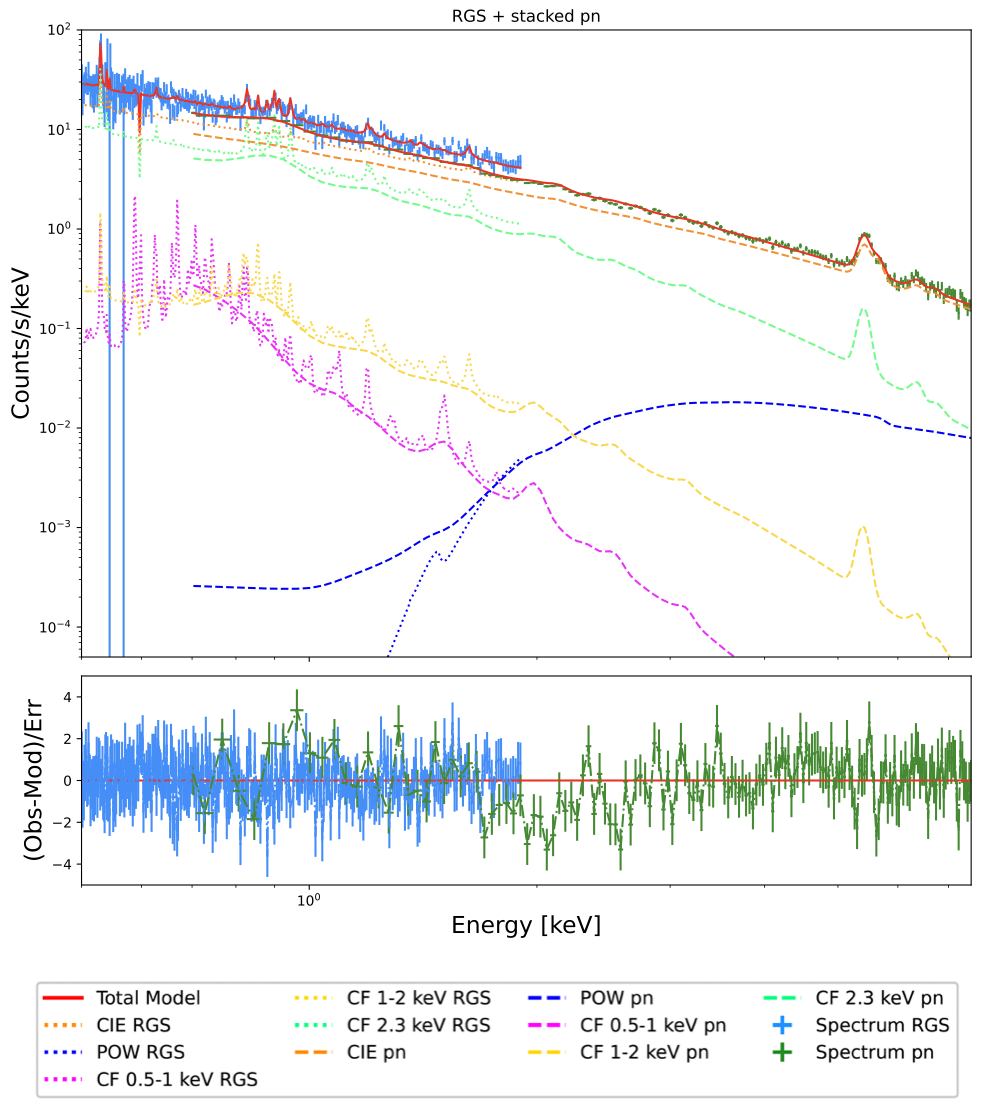}}
   \resizebox{!}{0.55\textwidth}{\includegraphics[scale=0.3]{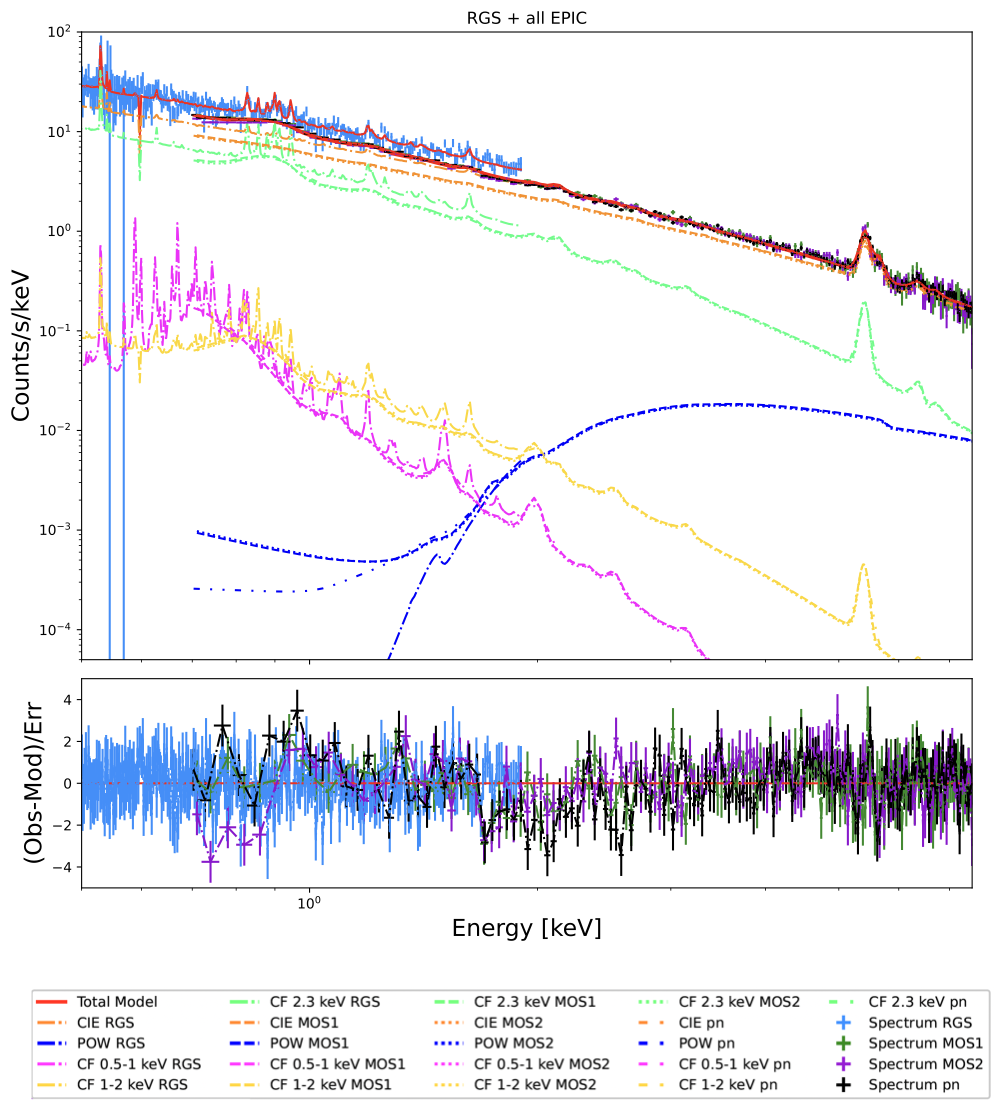}}
   \caption{Spectral fit of RGS + stacked MOS1 (light blue and green points, upper left panel), RGS + stacked MOS2 (light blue and green points, upper right panel), RGS + stacked pn (light blue and green points, lower left panel), and RGS + all stacked EPIC (light blue, green, violet and black points, lower right panel). Residuals are shown in the bottom panel. 
   The model used for fitting is the combination of \emph{cie} and the isobaric CF split into three temperature intervals. The model components are represented with different colors and line styles in the plots. Orange lines represent the \emph{cie} component, and blue lines represent the power law component describing the ICM and AGN, respectively. Magenta, gold, and light green lines represent the three temperature bins (0.5-1, 1-2, and 2-3 keV) components.  
   }
    \label{fig:ObsID spectral fitting 2}
\end{figure*}

In this section we present the best-fit model plots of the RGS and EPIC joint fit. The spectra and corresponding best-fit models are shown in Fig. \ref{fig:ObsID spectral fitting 2}, where we also display each model component: the power-law and \emph{cie} components representing the ICM and AGN, as well as the three temperature bins of the CF component. The detailed fitting results are provided in Table \ref{tab:EPIC RGS joint fit}.

As stated in the previous section, a visual inspection reveals some residuals below 1-2 keV for ObsID 0900280101 and 0900280201 EPIC/MOS and EPIC/pn spectra. We suggested that these features may be attributed to systematic uncertainties in the calibration of EPIC detectors, which may account for 5-10\%. If we add also the RGS spectrum, those uncertainties may increase, with another source of discrepancy potentially related to differences in the atomic database which are on the order of 20\% \citep{2015Pinto, 2017DePlaa, 2018Pinto}. Therefore, the adopted model may not fully capture all the features observed in the spectra, particularly below 1 keV. 
To further address this issue, we also calculate the residuals in percentages below 2 keV. We found that residual values are consistent with 10\%.


\end{appendix}

\end{document}